\documentclass[10pt,a4paper]{article}

\usepackage[utf8]{inputenc}
\usepackage[T1]{fontenc}
\usepackage{authblk}
\usepackage{lmodern}
\usepackage{microtype}
\usepackage[a4paper,top=20mm,bottom=23mm,left=22mm,right=22mm,headheight=15pt,footskip=11mm]{geometry}
\usepackage{graphicx}
\usepackage[dvipsnames]{xcolor}
\usepackage{fancyhdr}
\usepackage{titlesec}
\usepackage{needspace}
\usepackage{amsmath,amssymb}
\usepackage[inline,shortlabels]{enumitem}
\usepackage{booktabs}
\usepackage{array}
\usepackage{multirow}
\usepackage{tabularx}
\usepackage{pifont}
\usepackage{listings}
\usepackage[most]{tcolorbox}
\usepackage{varwidth}
\usepackage[tableposition=top]{caption}
\usepackage{hyperref}
\hypersetup{colorlinks=true,citecolor=blue,linkcolor=blue,urlcolor=blue}
\usepackage[numbers]{natbib}
\usepackage{soul}

\usepackage{bm}
\usepackage{relsize}

\definecolor{QSTInk}{RGB}{31,40,54}
\definecolor{QSTNavy}{RGB}{24,55,91}
\definecolor{QSTMuted}{RGB}{102,116,135}
\definecolor{QSTPanelBlueDark}{RGB}{21,64,107}
\definecolor{QSTPanelBlueFrame}{RGB}{92,139,181}
\definecolor{QudeBlue}{RGB}{34,55,199}
\definecolor{QudeViolet}{RGB}{108,49,225}

\definecolor{QITLeanDecl}{RGB}{15,77,151}
\definecolor{QITLeanNamespace}{RGB}{16,108,137}
\definecolor{QITLeanProof}{RGB}{49,118,89}
\definecolor{QITLeanType}{RGB}{36,97,153}
\definecolor{QITLeanDomain}{RGB}{38,83,128}
\definecolor{QITLeanComment}{RGB}{94,111,124}
\definecolor{QITLeanString}{RGB}{26,118,116}
\definecolor{QITLeanWarn}{RGB}{145,69,60}

\DeclareRobustCommand{\LeanQuantumAlg}{\textnormal{\textsc{Lean-QuantumAlg}}}
\DeclareRobustCommand{\PaperShortTitle}{\LeanQuantumAlg}
\newcommand{\QudeLeapLogo}[1]{\includegraphics[#1]{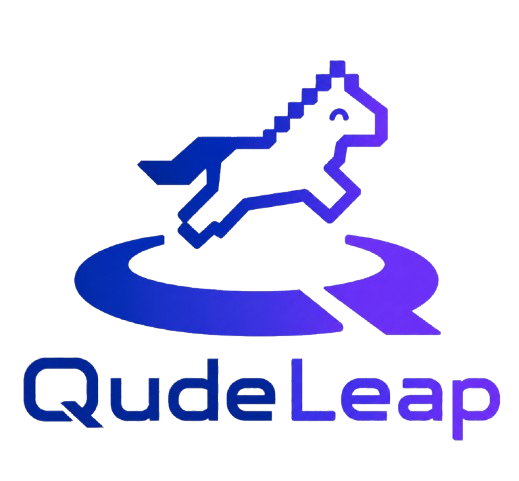}}

\titleformat{\section}
  {\Large\sffamily\bfseries\color{QSTPanelBlueDark}}
  {\thesection}{0.65em}{}
  [\vspace{0.15em}\color{QSTPanelBlueFrame!55}\titlerule]
\titleformat{\subsection}
  {\large\sffamily\bfseries\color{QSTNavy}}
  {\thesubsection}{0.65em}{}
\titleformat{\subsubsection}
  {\normalsize\sffamily\bfseries\color{QSTInk}}
  {\thesubsubsection}{0.65em}{}

\makeatletter
\renewcommand{\maketitle}{\thispagestyle{empty}\begingroup
  \setlength{\parindent}{0pt}\noindent
  \begin{minipage}[t]{0.8\textwidth}
    \vspace{0pt}{\raggedright\hyphenpenalty=10000\exhyphenpenalty=10000
      \sffamily\bfseries\color{QSTInk}\fontsize{23.5}{27.5}\selectfont
      \@title\par}\end{minipage}\hfill
  \begin{minipage}[t]{0.18\textwidth}
    \vspace{0pt}\raggedleft
    \QudeLeapLogo{width=2.3cm}\end{minipage}
  \vspace{4mm}

  \noindent
  {\color{QudeBlue}\rule{0.22\textwidth}{1.8pt}}{\color{QudeViolet}\rule{0.78\textwidth}{1.8pt}}\par
  \vspace{5mm}

  \begin{center}
    {\sffamily\normalsize\color{QSTInk}\@author\par}\end{center}
  \vspace{3mm}
  \@thanks
  \endgroup
  \setcounter{footnote}{0}}
\makeatother

\lstdefinelanguage{Lean4}{
  sensitive=true,
  alsoletter={_},
  morekeywords=[1]{
    abbrev,class,def,deriving,example,inductive,instance,lemma,
    noncomputable,opaque,structure,theorem
  },
  morekeywords=[2]{
    end,export,import,module,namespace,open,private,protected,public,section,
    variable,variables,where
  },
  morekeywords=[3]{
    by,calc,change,do,else,exact,exists,for,forall,from,fun,have,if,
    inferInstance,infix,infixl,infixr,intro,intros,let,letI,macro,match,
    obtain,rcases,refine,rfl,rw,show,syntax,termination_by,then,with
  },
  morekeywords=[4]{
    Bool,False,Fin,Int,List,Nat,None,Option,Prop,Real,Set,Some,Sort,
    String,True,Type,ENNReal,PMF
  },
  morekeywords=[5]{axiom,sorry,admit},
  morekeywords=[6]{
    Algorithm,AlgorithmRun,Circuit,CShape,ExecutionModel,
    ExecutionState,GateModel,OracleModel,Output,Problem,ProbabilityLowerBound,
    QuantumLayout,Register,ResourceBudget,ResourcePool,RuntimeCtx,Solves,
    SpacePool,StateRequirement,UnitaryLeaf,UnitaryOperator
  },
  morecomment=[l]{--},
  morecomment=[s]{/-}{-/},
  morestring=[b]",
}

\lstdefinestyle{QudeLeanListing}{
  language=Lean4,
  basicstyle=\fontsize{9.8}{11.5}\selectfont\ttfamily\color{QSTInk},
  keywordstyle=[1]\color{QITLeanDecl}\bfseries,
  keywordstyle=[2]\color{QITLeanNamespace}\bfseries,
  keywordstyle=[3]\color{QITLeanProof},
  keywordstyle=[4]\color{QITLeanType}\bfseries,
  keywordstyle=[5]\color{QITLeanWarn}\bfseries,
  keywordstyle=[6]\color{QITLeanDomain}\bfseries,
  commentstyle=\color{QITLeanComment},
  stringstyle=\color{QITLeanString},
  numbers=none,
  xleftmargin=0pt,
  framexleftmargin=0pt,
  columns=fullflexible,
  keepspaces=true,
  showstringspaces=false,
  tabsize=2,
  breaklines=true,
  breakatwhitespace=false,
  breakindent=1.15em,
  breakautoindent=true,
  aboveskip=0pt,
  belowskip=0pt,
  literate=
    {_}{{\textunderscore\allowbreak}}1
    {∀}{{\ensuremath{\forall}}}1
    {∃}{{\ensuremath{\exists}}}1
    {→}{{\ensuremath{\to}}}1
    {←}{{\ensuremath{\leftarrow}}}1
    {↔}{{\ensuremath{\leftrightarrow}}}1
    {∧}{{\ensuremath{\wedge}}}1
    {∨}{{\ensuremath{\vee}}}1
    {¬}{{\ensuremath{\neg}}}1
    {≤}{{\ensuremath{\leq}}}1
    {≥}{{\ensuremath{\geq}}}1
    {≠}{{\ensuremath{\neq}}}1
    {∈}{{\ensuremath{\in}}}1
    {⟨}{{\ensuremath{\langle}}}1
    {⟩}{{\ensuremath{\rangle}}}1
    {λ}{{\ensuremath{\lambda}}}1
    {α}{{\ensuremath{\alpha}}}1
    {β}{{\ensuremath{\beta}}}1
    {θ}{{\ensuremath{\theta}}}1
    {η}{{\ensuremath{\eta}}}1
    {ψ}{{\ensuremath{\psi}}}1
    {ρ}{{\ensuremath{\rho}}}1
    {ℕ}{{\ensuremath{\mathbb{N}}}}1
    {ℤ}{{\ensuremath{\mathbb{Z}}}}1
    {ℚ}{{\ensuremath{\mathbb{Q}}}}1
    {ℝ}{{\ensuremath{\mathbb{R}}}}1
    {×}{{\ensuremath{\times}}}1
    {ˣ}{{\ensuremath{^{\times}}}}1
    {⁻¹}{{\ensuremath{^{-1}}}}2
    {‖}{{\ensuremath{\|}}}1
    {∑}{{\ensuremath{\sum}}}1
    {·}{{\ensuremath{\cdot}}}1
    {⦃}{{\ensuremath{\{\!\!\{}}}1
    {⦄}{{\ensuremath{\}\!\!\}}}}1
    {∘}{{\ensuremath{\circ}}}1
    {⊗}{{\ensuremath{\otimes}}}1
    {ε}{{\ensuremath{\varepsilon}}}1
    {π}{{\ensuremath{\pi}}}1
}

\newtcblisting[auto counter]{leanlisting}[3][]{enhanced,
  breakable,
  listing only,
  listing engine=listings,
  listing options={style=QudeLeanListing},
  colback=gray!5,
  colframe=gray!55,
  boxrule=0.4pt,
  arc=0.6mm,
  left=4pt,
  right=4pt,
  top=4pt,
  bottom=4pt,
  title={Listing~\thetcbcounter: #2},
  fonttitle=\sffamily\small,
  coltitle=black,
  attach boxed title to top left={yshift=-2mm,xshift=4mm},
  varwidth boxed title=\dimexpr\linewidth-8mm\relax,
  boxed title style={colback=gray!18,sharp corners,boxrule=0pt,boxsep=2pt},
  label={#3},
  before skip=0.6\baselineskip,
  after skip=0.65\baselineskip,
  #1
}

\newtcbinputlisting[use counter from=leanlisting]{\leaninputlisting}[3]{enhanced,
  breakable,
  listing only,
  listing engine=listings,
  listing file={#3},
  listing options={style=QudeLeanListing},
  colback=gray!5,
  colframe=gray!55,
  boxrule=0.4pt,
  arc=0.6mm,
  left=4pt,
  right=4pt,
  top=4pt,
  bottom=4pt,
  title={Listing~\thetcbcounter: #1},
  title after break={Listing~\thetcbcounter\ (continued)},
  fonttitle=\sffamily\small,
  coltitle=black,
  attach boxed title to top left={yshift=-2mm,xshift=4mm},
  varwidth boxed title=\dimexpr\linewidth-8mm\relax,
  boxed title style={colback=gray!18,sharp corners,boxrule=0pt,boxsep=2pt},
  label={#2},
  before skip=0.6\baselineskip,
  after skip=0.65\baselineskip
}

\newcommand{\nc}{\newcommand}

\nc{\bra}[1]{\langle#1|}
\nc{\ket}[1]{|#1\rangle}
\nc{\braket}[2]{\langle#1|#2\rangle}

\nc{\lean}[1]{\texttt{#1}}
\nc{\leanpath}[1]{\texttt{#1}}

\nc{\cmark}{\ding{51}}
\nc{\pmark}{\textsf{\footnotesize partial}}
\nc{\xmark}{{\color{darkgray} \ding{55}}}

\nc{\specialcell}[2][c]{\begin{tabular}[#1]{@{}c@{}}#2\end{tabular}}

\title{\fontsize{18.2}{21}\selectfont
  Lean-QuantumAlg: \\
  A Framework for AI Agents to Produce \\
  Quantum Algorithms that Humans Can Trust}

\author[1,2]{Lei Zhang\thanks{leizhang116.4@gmail.com}}
\author[1,2]{Mingrui Jing}
\author[1]{Yusheng Zhao}
\author[2]{Jiahong Liu}
\author[2]{Xin Wang\thanks{felixxinwang@hkust-gz.edu.cn}}
\affil[1]{\small QudeLeap Research, Shanghai 200030, China}
\affil[2]{\small The Hong Kong University of Science and Technology (Guangzhou), Guangdong 511453, China}
\date{}

\ifdefined\hypersetup
\hypersetup{pdfcreator={},pdfproducer={}}
\else
\ifdefined{/Creator () /Producer ()}\fi
\fi
\ifdefined\pdfinfoomitdate\pdfinfoomitdate=1\fi
\ifdefined\pdfsuppressptexinfo\pdfsuppressptexinfo=-1\fi
\ifdefined\pdftrailerid\pdftrailerid{}\fi
\begin{document}
\maketitle

\vspace{-5mm}
\begin{center}
  \small\sffamily\color{QSTMuted}\raisebox{-0.5ex}{\QudeLeapLogo{height=3.2mm}}\hspace{2pt}\href{https://github.com/QudeLeap/Lean-QuantumAlg}{github.com/QudeLeap/Lean-QuantumAlg}\end{center}

\begin{abstract}
AI agents are increasingly capable of constructing quantum algorithms and proofs, yet machine-checked proofs alone cannot guarantee that the intended problem is solved or that resource claims match the verified computation.
Here we introduce \LeanQuantumAlg{}, a Lean~4 framework for trustworthy AI-assisted development of quantum algorithms.
Its Algorithm--Solves--Problem contract binds correctness, success probability, borrowed-state restoration and quantum and classical resource bounds to a single execution, while separating the guarantees that Lean checks from the problem specifications and execution models that humans review.
We demonstrate the framework on nine AI-formalized quantum algorithms, from Grover search to quantum attacks on RSA-2048 and P-256, with less than 1\% of each algorithm's imported source code requiring statement-level human review once the shared framework has been audited.
By integrating reusable mathematical foundations, certified algorithmic primitives and execution-grounded verification, \LeanQuantumAlg{} provides a basis for AI agents to construct and certify quantum algorithms, opening a path toward scalable AI-driven discovery of quantum algorithms.
\end{abstract}

\section{Introduction}
\label{sec:introduction}

Quantum algorithms combine operations on quantum systems with classical computation to solve computational problems.
Their promise includes quadratic query savings for unstructured search~\cite{grover1996fast} and polynomial-time algorithms for integer factorization~\cite{shor1997polynomial}.
The corresponding algorithm statements combine two kinds of guarantees: correctness and resources.
A correctness guarantee states that the algorithm solves the intended problem under the stated input conditions with the claimed success probability; resource guarantees bound the quantum and classical operations and storage used to do so.

Artificial intelligence (AI) has begun to assist quantum algorithm development through circuit optimization~\cite{ruiz2025alphatensorq} and proposals tested on quantum hardware~\cite{deepmind2026alphaevolve}.
In mathematics, researchers can work with AI agents that use software tools to check their work~\cite{tsoukalas2026nexus}.
Such checking relies on formalization, in which the relevant definitions, statements and proofs are expressed in the programming language of a proof assistant.
A proof assistant, such as Lean~\cite{demoura2021lean}, is a software system that supports this process and checks whether a proof establishes the stated proposition.
Recent AI results include proofs of open Erd\H{o}s problems~\cite{tsoukalas2026nexus}, of a version of the Navier--Stokes problem~\cite{openai2026navierstokes} and of Fermat's Last Theorem~\cite{anthropic2026flt}, as well as manuscripts on hundreds of mathematical problems~\cite{openai2026mathrepo}.
These results suggest that AI agents are capable of producing quantum algorithms accompanied by proofs of their correctness and resource bounds.

While formalization lets a proof assistant check results produced by AI, whether the formal statements express the intended claims remains a subject of human review, since changing the meaning of a formal statement changes the result established.
For instance, an ambiguous Erd\H{o}s problem required a second proof after its intended reading was clarified~\cite{sothanaphan2026erdos728}, and an audit of Lean benchmarks identified vacuous statements and incorrect translations of informal problems~\cite{ammanamanchi2026faults}.
If a formalization is presented as ``checking a written argument'', the correspondence between the two also matters: a different proof of the same statement still establishes that statement, but does not by itself validate the original argument.
An analysis of the Navier--Stokes work found differences between its Lean proof and the released natural-language argument~\cite{bastounis2026navierstokes}.
Organizing this remaining human review therefore becomes increasingly important~\cite{tao2026ageofai} as AI expands theoretical production.

\begin{figure}[t]
  \centering
  \includegraphics[width=\linewidth]{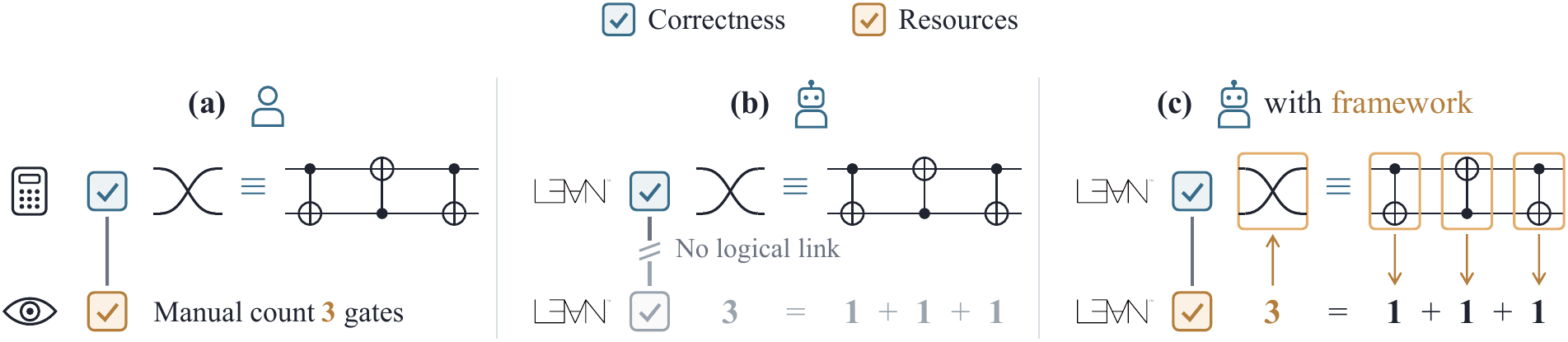}
  \caption{A schematic illustration of three ways to prove correctness and resources of the statement ``a SWAP gate can be implemented by 3 CNOT gates'', by a human, artificial intelligence (AI), and AI with a proof framework, respectively. Here ``correctness'' refers to the equivalence of implementation, and ``resources'' refers to the cost of implementation. \textbf{(a)}~The statement is proved by exact matrix calculation and manual resource counting. \textbf{(b)}~The statement is claimed to be proved by separately verifying the correctness and resources, or \textbf{(c)}~within a framework the statement is proved by additionally linking the resource unit with a CNOT gate.}
  \label{fig:swap}
\end{figure}

For quantum algorithms, this review concerns several connected parts of an algorithm statement, because an algorithm's name alone does not fix the execution model, which specifies the allowed gates, oracle queries and classical operations.
Without human review, AI agents can make mistakes in how the correctness and resource guarantees refer to the same implementation in that model.
As a simple example, consider the statement ``a SWAP gate can be implemented by 3 controlled-NOT (CNOT) gates''~\cite{nielsen2010quantum}.
An AI agent may account for the $3$ CNOT gates by proving $3 = 1 + 1 + 1$, a count disconnected from the circuit whose correctness is proved, as illustrated in Figure~\ref{fig:swap}.
Larger algorithms leave room for two further mistakes.
A resource count can leave out costly work: an expensive subroutine can be declared a single primitive gate, or the repeated runs and classical post-processing that recover the answer, such as the continued-fraction expansion in order finding~\cite{shor1997polynomial}, can be omitted from the count.
A borrowed qubit, which may be entangled with other qubits and must be handed back unchanged~\cite{gidney2017factoring}, can also be returned in the correct local state while its entanglement with those qubits is destroyed.

Current formal tools address part of this review problem.
Several of them prove that quantum programs meet their specifications, the mathematical requirements that an implementation is meant to satisfy~\cite{chareton2021automated,hietala2021proving,rand2018qwire,zhou2023coqq,ren2026qedia}.
In particular, Qbricks~\cite{chareton2021automated} supports deductive reasoning about circuit families, including correctness and resource bounds, and Ref.~\cite{peng2023formally} establishes correctness and gate-count bounds for the same implementation of Shor's algorithm~\cite{shor1997polynomial}.
Type-based analyses derive circuit-width bounds~\cite{colledan2025circuit} and extend to gate count and depth~\cite{colledan2025flexible}, while languages and libraries such as Quipper~\cite{green2013quipper} and Qualtran~\cite{harrigan2024qualtran} support the construction and resource estimation of large quantum algorithms.
Research on autoformalization, the automated translation of mathematical statements and proofs into formal language, addresses the review of mathematical statements: it evaluates whether formal statements preserve their intended meaning~\cite{lu2025formalalign,zhang2026beyond} and develops workflows intended to preserve correspondence to the written argument~\cite{pegden2026trellis}.
However, none of these tools ties correctness and resource guarantees to one complete execution: most address only one of the two, and those that establish both bound only the gates and qubits of the quantum circuit, leaving repetitions and classical processing outside the bounds.
Connecting these partial guarantees is left to the reviewer, so for a development produced by AI agents, the part that requires human review is not delimited.
How AI agents can produce algorithm statements that unify correctness and resource guarantees remains an open problem.

We present \LeanQuantumAlg{}, a Lean~4 package built on Mathlib~\cite{mathlib} that provides a formalization framework implementing an Algorithm--Solves--Problem contract.
A \lean{Problem} states all problem settings independently of a candidate implementation, an \lean{Algorithm} supplies a candidate implementation, and the framework derives the relevant probabilities, states and resource use from its execution.
The semantic relation \lean{Solves} requires that this algorithm satisfy the problem, restore borrowed systems and meet the resource bounds, with every guarantee referring to the same execution.
A resource count for another circuit thus cannot discharge these bounds.
The execution model owns every operation that can be executed or charged, and the repeated runs and classical post-processing of the execution are counted with its quantum gates.
Restoration compares each borrowed system jointly with an external reference, so returning only its local state does not satisfy it.
Within this framework, AI agents construct algorithms and proofs, and Lean checks the proof that an algorithm solving the problem exists.
Human review therefore focuses on the problem and the execution model, including which operations the model grants as primitives and whether the problem requires the final answer, since Lean checks that AI agents have proved exactly the reviewed statement.
This organization makes the specification choices visible and enforces the connections among the guarantees.
To demonstrate this division of work, we use AI agents to prove nine quantum algorithms.
We show that, once the shared framework has been audited, the code requiring human review accounts for less than 1\% of the lines of code in the import closure of every algorithm, as shown in Table~\ref{tab:footprint}, providing a basis for scaling the construction and formal verification of quantum algorithms while keeping the subject of human review explicit.
\section{Main results}
\label{sec:main-results}

We first describe how the framework connects an algorithm to its claim and how this determines what a human reads.
Section~\ref{sec:overview} describes how \LeanQuantumAlg{} works and measures the code requiring human review for nine algorithms, with the measurements reported in Table~\ref{tab:footprint}.
Section~\ref{sec:comparison} compares the framework with the formal tools that certify quantum programs or bound their resources, summarizing their capabilities in Table~\ref{tab:comparison}; the cell-by-cell evidence for that table is in Appendix~\ref{appendix:evidence}.

\subsection{How the framework works}
\label{sec:overview}

\begin{figure}[ht]
  \centering
  \includegraphics[width=0.65\linewidth]{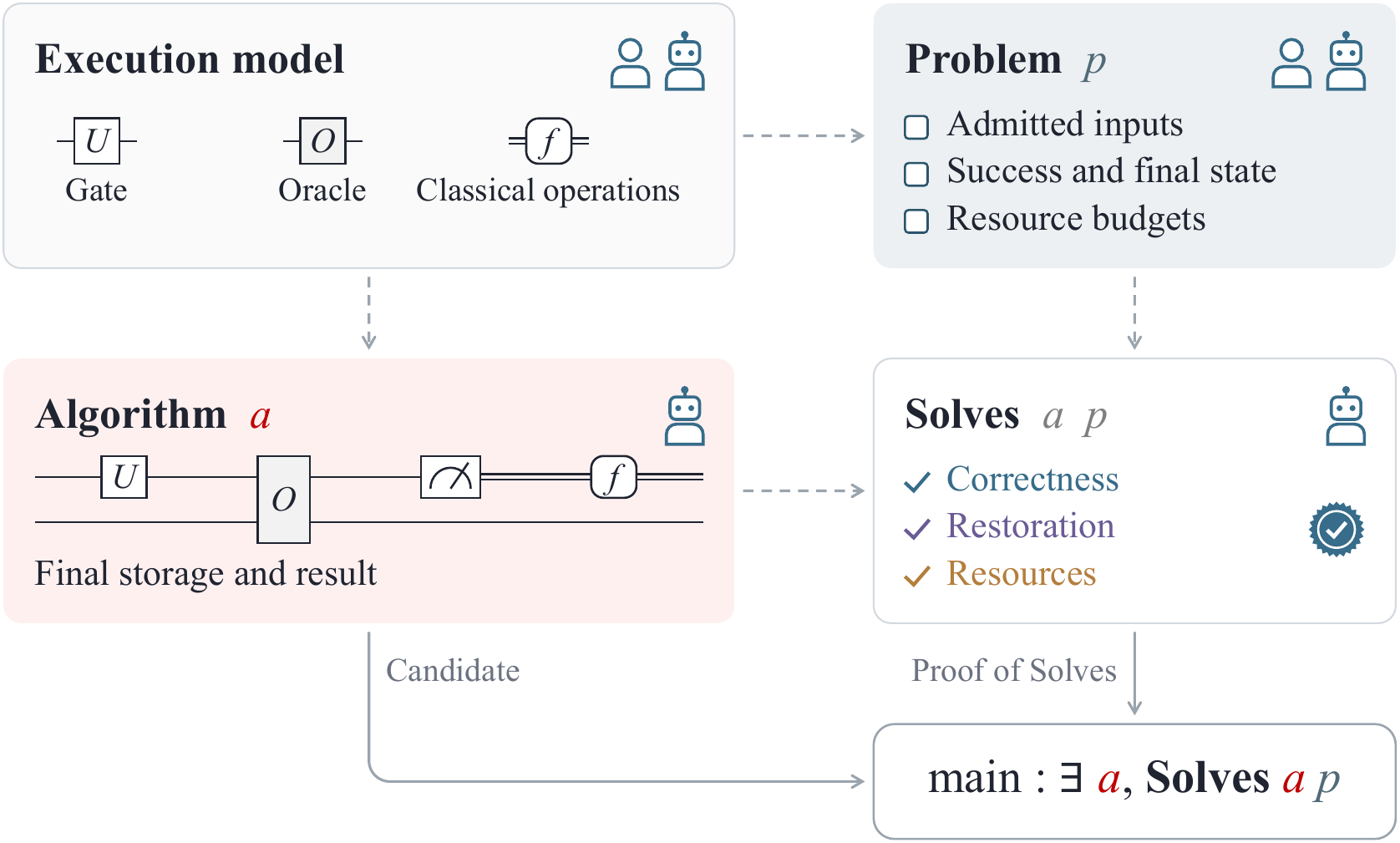}
  \caption{Relationships among the objects used to certify an algorithm.
    The execution model fixes the gates $U$, oracle queries $O$ and classical operations $f$ from which an algorithm's circuit is built.
    The model is an argument of both a stated \lean{Problem} $p$ and a candidate \lean{Algorithm} $a$, which are in turn the arguments of \lean{Solves}, as the dashed arrows indicate.
    Solid arrows show how the candidate and a proof of \lean{Solves a p} establish the existential theorem \lean{main}.
    Person icons mark what a human reviews, and robot icons mark the work of AI agents.}
  \label{fig:overview}
\end{figure}

As shown in Figure~\ref{fig:overview}, a stated problem is connected to a candidate implementation through the relation \lean{Solves}.
An \emph{execution model} fixes what a program may do.
It owns the tokens for native gates, oracle queries and classical operations, each token determining its resource event and its action in the supplied environment, and nothing outside the model can be executed or charged.
A \lean{Problem} allows the claim to be stated independently of an implementation.
An \lean{Algorithm} contains a circuit in the model together with a declaration of its final storage and returned result.
The framework computes its behavior from this execution and its resource use from the same circuit and execution paths.
The relation \lean{Solves} connects the two through obligations that all refer to the same execution of that run.
Every executable algorithm in the library ends in a final theorem \lean{main} stating that some \lean{Algorithm} solves its \lean{Problem}.

\lean{Solves} guarantees that for each allowed instance of the stated problem, every input it admits and every external reference system correlated with that input, one execution of the algorithm accepts with at least the required probability, leaves the final state as the problem demands, restores every borrowed system jointly with the reference, and on every path of positive probability stays within the declared event, space and quantum-time budgets.
The four verbs refer to the same execution, and there is no second object whose cost is reported.
The guarantee is relative to the execution model named in the theorem.
The existential \lean{main} certifies that a solving algorithm exists, and its proof exhibits a solving candidate.

The design fixes who reads what.
The human reads the statement of \lean{main}, the \lean{Problem} it names and the execution model, and, once for the whole framework, the shared definitions that give these words their meaning.
AI agents construct the \lean{Algorithm} and prove that it satisfies \lean{Solves}.
Lean checks that the proofs are valid and therefore that the exported algorithm meets every obligation of \lean{Solves}.
Once the problem and its defining expressions have been reviewed, \lean{Solves} lets the reviewer leave the candidate construction and proof to Lean; Section~\ref{sec:review} works this out in detail.

How much a reviewer reads depends on which parts have already been reviewed; Lean and Mathlib are trusted throughout.
If the framework has yet to be audited, the reviewer reads every framework definition on which the statement of \lean{main} depends; Table~\ref{tab:footprint} labels this column ``Framework untrusted''.
If the shared semantics are already trusted, including the execution model, \lean{Problem}, \lean{Algorithm}, \lean{Solves} and the general-purpose mathematics beneath them, the reviewer reads the reusable components and certified primitives used by the algorithm together with the algorithm's own modules; this is the column ``Core and Util trusted''.
If the whole framework is trusted, the reviewer reads only the definitions in the algorithm's own modules and the statement of \lean{main}; this is the column ``Framework trusted'', representing review of one agent-written development on top of an audited framework.

\begin{table}[t]
  \centering
  \caption{A summary of the nine quantum algorithms that AI agents proved within \LeanQuantumAlg{}.
  For each algorithm, the table gives the percentage of code requiring human review, the number of theorems on which the proof of its final theorem depends and the lines of code (LOC) in its import closure.
  Here, the three percentages correspond to a reviewer who trusts the whole framework, only its Core and Util modules, or none of the framework.}
  \label{tab:footprint}
  \begin{tabular}{@{}lrrrrr@{}}
    \toprule
    Algorithm & \multicolumn{3}{c}{Code requiring human review (\%)} & Theorems & LOC \\
    \cmidrule(lr){2-4}
     & \specialcell{Framework\\trusted} & \specialcell{Core and Util\\trusted} & \specialcell{Framework\\untrusted} & & \\
    \midrule
    Deutsch--Jozsa~\cite{deutsch1992rapid,dewolf2019lecturenotes} & 0.17 & 0.97 & 11.86 & 785 & 25\,858 \\
    Bernstein--Vazirani~\cite{bernstein1997quantum,dewolf2019lecturenotes} & 0.13 & 0.94 & 11.85 & 791 & 25\,808 \\
    Grover search~\cite{grover1996fast,dewolf2019lecturenotes} & 0.21 & 0.88 & 8.38 & 1\,101 & 37\,742 \\
    Quantum phase estimation~\cite{kitaev1995quantum,lin2022qasc} & 0.29 & 0.99 & 10.72 & 878 & 29\,374 \\
    Amplitude estimation~\cite{brassard2002quantum,lin2022qasc} & 0.24 & 0.86 & 7.13 & 1\,408 & 45\,087 \\
    Order finding~\cite{shor1997polynomial} & 0.29 & 0.90 & 6.23 & 1\,582 & 54\,070 \\
    Finite-cyclic discrete logarithm~\cite{shor1997polynomial,shukla2024uniformsuperposition} & 0.72 & 2.05 & 6.24 & 2\,469 & 82\,505 \\
    Quantum attack on RSA-2048~\cite{ekera2017quantum,ekera2023shortdlp} & 0.43 & 0.73 & 2.29 & 6\,439 & 192\,975 \\
    Quantum attack on P-256~\cite{proos2004shor,roetteler2017quantum,nist2023sp800186} & 0.09 & 0.43 & 1.41 & 6\,674 & 309\,481 \\
    \bottomrule
  \end{tabular}
\end{table}

Table~\ref{tab:footprint} reports the source lines requiring human review under these three conditions for the nine algorithms shipped with the framework, from Deutsch--Jozsa to the quantum attacks on RSA-2048 and P-256.
All nine final theorems have the existential form described above, and the measurements use the same library version as the listings.
The measure is taken at the level of declarations.
The \emph{statement dependencies} of \lean{main} are the library definitions on which its statement depends, directly or indirectly, through types and definition bodies; proof bodies are excluded.
When traversal reaches an \lean{Algorithm} definition, it follows its type and stops before its body; Lean checks the proof that the implementation satisfies \lean{Solves}.
The fraction of code requiring human review counts the source lines of these dependencies and of the statement of \lean{main} outside the trusted modules, relative to the lines of code (LOC) in the import closure of the algorithm.
Proofs, the statements of intermediate lemmas and the construction of the circuit do not enter the review count.
The denominator includes every library module in the import closure, including definitions and proofs that the final theorem does not use.
The percentages measure source lines, rather than review time or the difficulty of understanding them.
Appendix~\ref{appendix:footprint} states the rules.

Once the framework has been audited, the first percentage column is below 1\% for every algorithm, with a maximum of approximately 0.72\% for the finite-cyclic discrete logarithm.
The six cases from Deutsch--Jozsa to order finding each require less than 0.3\%.
With only the shared semantics trusted, the share ranges from approximately 0.43\% to 2.05\%.
The ``Framework untrusted'' column includes the shared definitions reached from each final statement, giving shares from approximately 1.41\% to 11.86\%.
These definitions can be reviewed once and reused across algorithms; the difference between the first and third percentage columns shows how much of each statement's source-line count is shared framework code.

\subsection{Comparison with existing work}
\label{sec:comparison}

\begin{table}[t]
  \centering
  \small
  \caption{
    A comparison of some packages for the formal verification or resource analysis of quantum algorithms, according to the correctness and resource metrics they support (\cmark{}), support with restrictions (\pmark{}), or do not support (\xmark{}).
    Here, dynamic circuit refers to whether a package supports mid-circuit measurement with classical feedback inside the certified program, and same object refers to whether it proves the correctness and resource bounds for one object.
    In the Depth/Time column, depth refers to a bound on the depth of a circuit, and time to a bound on the quantum computation time of every complete execution path of an algorithm.
    }
  \label{tab:comparison}
  \resizebox{\linewidth}{!}{\begin{tabular}{@{}lccccccccc@{}}
    \toprule
    & \multicolumn{3}{c}{Supported correctness} & \multirow{3}{*}{\specialcell{Same\\object}}
      & \multicolumn{5}{c}{Supported resources} \\
    \cmidrule(lr){2-4}\cmidrule(lr){6-10}
    Package & \specialcell{Expected\\output} & \specialcell{Success\\probability}
      & \specialcell{Dynamic\\circuit} &
      & Gates & Qubits & \specialcell{Depth/\\Time}
      & \specialcell{Classical\\bits} & \specialcell{Classical\\operations} \\
    \midrule
    Qbricks~\cite{chareton2021automated}
      & \cmark & \cmark & \xmark & \cmark & \cmark & \cmark & \xmark & \xmark & \xmark \\
    SQIR~\cite{hietala2021verified,hietala2021proving,peng2023formally}
      & \cmark & \cmark & \cmark & \cmark & \cmark & \cmark & \xmark & \xmark & \xmark \\
    QWIRE~\cite{paykin2017qwire,rand2018qwire}
      & \cmark & \cmark & \cmark & \xmark & \xmark & \pmark & \xmark & \xmark & \xmark \\
    CoqQ~\cite{zhou2023coqq}
      & \cmark & \cmark & \cmark & \xmark & \xmark & \pmark & \xmark & \xmark & \xmark \\
    QEDIA~\cite{ren2026qedia}
      & \cmark & \cmark & \cmark & \xmark & \xmark & \pmark & \xmark & \xmark & \xmark \\
    QuRA~\cite{colledan2025flexible,colledan2025circuit}
      & \xmark & \xmark & \pmark & \xmark & \cmark & \cmark & \cmark & \cmark & \pmark \\
    \midrule
    Lean-QuantumAlg
      & \cmark & \cmark & \cmark & \cmark & \cmark & \cmark & \cmark & \cmark & \cmark \\
    \bottomrule
  \end{tabular}}
\end{table}

Formal tools for quantum programs can be broadly divided into correctness verification, which proves that a program meets its specification, and resource analysis, which bounds the resources of the circuits that a program generates.
Table~\ref{tab:comparison} compares representative tools of both classes in the versions described by their cited papers, and Appendix~\ref{appendix:evidence} gives the evidence for individual cells.

Among correctness verifiers, Qbricks~\cite{chareton2021automated} and SQIR~\cite{hietala2021verified,hietala2021proving} already bind correctness and some resource bounds to one implementation.
Qbricks combines functional, probabilistic and size specifications for one circuit term in a Hoare-style logic, and Ref.~\cite{peng2023formally} uses SQIR to certify an end-to-end implementation of Shor's algorithm with a gate-count bound for the circuit whose success probability it establishes.
In both, however, the certified resources are the gate and qubit counts of a single circuit: oracles enter as concrete circuits whose gates are counted, and neither the repetitions of the circuit nor the classical post-processing of Shor's algorithm receives a cost bound.
Qbricks also excludes measurement from its language, whereas full SQIR can branch on measurement outcomes~\cite{hietala2021proving}.

Other verifiers address programs in which measurement outcomes control later computation.
QWIRE~\cite{paykin2017qwire,rand2018qwire} supports such programs through dynamic lifting, CoqQ~\cite{zhou2023coqq} verifies conditionals and loops that branch on measurements with a quantum Hoare logic, and QEDIA~\cite{ren2026qedia} formalizes a quantum Hoare logic in Lean for programs with classical variables.
QEDIA also reports constructing its library and proofs with AI agents under human supervision.
The types of these tools fix the dimensions of the quantum systems, but the cited versions establish no resource bounds for the verified programs.

Resource analysis instead starts from the quantities to be bounded.
QuRA derives circuit-width bounds by effect typing~\cite{colledan2025circuit}, and Ref.~\cite{colledan2025flexible} extends its analysis to parametric bounds on width, gate count and depth with a soundness proof; its Depth/Time mark records this proof for depth, because the tool's implementation of local metrics such as depth was still in progress.
These bounds concern the generated circuit: whether it implements the intended algorithm needs a separate correctness argument, and the classical computation in the host that builds the circuit lies outside them.

Our work brings both goals into one algorithm statement, summarized in the last row of Table~\ref{tab:comparison}.
The correctness, final-state, borrowed-restoration and resource obligations of \lean{Solves} all refer to one complete execution, including measurement-dependent control, repeated quantum work and declared classical operations.
Gates and oracle queries are separate event coordinates, and space is counted as live quantum systems of each dimension.
Its Depth/Time mark refers to quantum computation time, which \lean{Solves} bounds on every complete execution path; general circuit depth remains a separate analysis at the \lean{Circuit} level.
These guarantees are relative to the declared execution model, which a human reviews together with the problem.
\Needspace{9\baselineskip}
\section{The framework: problems, algorithms, and \lean{Solves}}
\label{sec:framework}

\subsection{Formal statements and proofs in Lean}
\label{sec:formal-statements}

A formal development begins by defining the objects about which it makes claims.
In Lean, a type describes a class of objects, and a structure collects named fields into one object.
For example, the \lean{Problem} structure below collects input conditions, success and state requirements, and resource bounds; a particular problem supplies a value for each field.
A proposition then states a mathematical claim about such objects.
Its meaning depends on the definitions it refers to, including the meaning of its inputs, operations and resource units.

To prove a proposition, a human or an automated tool constructs a proof that Lean can check against the proposition's exact statement.
Lean's kernel checks the resulting proof using the rules of its logic and the declared definitions and assumptions~\cite{demoura2021lean}.
Formal verification of an algorithm uses this process to establish that the implementation satisfies a specification expressed as a proposition.
The scientific interpretation of that specification remains a matter for human review: a valid proof establishes precisely what was stated.
Section~\ref{sec:review} describes this review and the standard checks on proof completion and assumptions.

The following listings use these ideas in three steps.
\lean{Problem} and \lean{Algorithm} describe the requirements and the implementation, respectively.
\lean{Solves} is a proposition relating them, with one field for each obligation that must be proved.
The final theorem \lean{main} states that an algorithm satisfying these obligations exists.
Its proof supplies an implementation and proves that it meets them, so a reader can understand the claim by following the statement and its definitions while Lean checks the construction and proof.

\LeanQuantumAlg{} is organized as one Lean library.
The framework proper lives under \leanpath{QuantumAlg/Core} (execution models, circuits, the interpreter, resource accounting and the three interface types of this section) and \leanpath{QuantumAlg/Util} (general-purpose mathematics).
The reusable building blocks shipped with it live under \leanpath{QuantumAlg/Components} (gate and oracle components) and \leanpath{QuantumAlg/Primitives} (certified subroutines such as the quantum Fourier transform, amplitude amplification and modular arithmetic), and the algorithms live under \leanpath{QuantumAlg/Algorithms}.
The listings reproduce the declarations with source comments omitted, and each block names its source file and original line range.
Explanations appear in the surrounding text; where only a theorem statement is shown, its proof is checked in the source development.

\subsection{\lean{Problem}: the claim, written once}
\label{sec:problem}

The \lean{Problem} record in Listing~\ref{lst:problem} is indexed by an execution model, an input context, an output context and an output shape.
Its type mentions no circuit or algorithm: the specification supplies the requirements against which a candidate implementation is certified.
This interface allows the requirements to be stated independently of a candidate.
The reviewer checks these requirements and their defining expressions against the intended claim; \lean{Solves} checks that the candidate meets them.

\begin{leanlisting}{The \lean{Problem} structure}{lst:problem}
-- source: QuantumAlg/Core/Algorithm/Problem.lean:49-69
public structure Problem (model : ExecutionModel) (pre post : RuntimeCtx model.kind)
    (outputShape : CShape) : Type 1 where
  Witness : Type
  valid : model.Environment → Witness → ExecutionState (Execution.Store pre) pre.register → Prop
  initialization : model.Environment → Witness →
    ExecutionState (Execution.Store pre × model.HiddenState) pre.register → Prop
  accepts : model.Environment → Witness → Execution.Store pre → outputShape.Value → Prop
  successLB : model.Environment → Witness → ProbabilityLowerBound
  states : model.Environment → Witness → List (StateRequirement pre post)
  resources : ResourceBudget model
  quantumTimeLimit : Nat
\end{leanlisting}

Each field makes one commitment of the claim explicit.
The \lean{Witness} field supplies data used only to state the specification, and the other variable fields receive both an environment and a value of that type before their ordinary arguments.
\lean{valid} is the input promise, a predicate of the environment, the witness and the local classical--quantum input state.
It sees only the marginal that the algorithm receives, so it cannot exclude an external reference system correlated with that input.
The environment supplies the interpretations of the model's declared quantum and classical oracles, while \lean{initialization} admits joint initial states of the caller's classical store and the hidden service state of the model.
\lean{accepts} relates the initial classical value to the returned result, so success is a relational event and correlated inputs remain meaningful.
\lean{successLB} is a probability threshold certified to be at most one.
\lean{states} lists requirements on the final state, which may be pure, mixed, marginal, joint or dirty.
Using the budgets defined in Listing~\ref{lst:budgets}, the \lean{resources} field bounds every native event coordinate and the peaks of resident and auxiliary space.
Finally, \lean{quantumTimeLimit} bounds the quantum computation time of the complete execution, under the fixed accounting convention of Section~\ref{sec:accounting}.

\begin{leanlisting}{Probability threshold and budgets}{lst:budgets}
-- source: QuantumAlg/Core/Algorithm/Problem.lean:33-47
public structure ProbabilityLowerBound where
  value : ENNReal
  le_one : value ≤ 1

public structure ResourceBudget (model : ExecutionModel) where
  events : ResourcePool model
  resident : SpacePool model.kind
  auxiliary : SpacePool model.kind
\end{leanlisting}

Quantum phase estimation illustrates the fields.
Its \lean{Witness} records an $n$-qubit unitary $U$, a borrowed eigenstate $\ket{\psi}$, and an eigenphase $\theta$ satisfying $U\ket{\psi} = e^{2\pi i\theta}\ket{\psi}$; the problem is indexed by a precision width $t$.
\lean{valid} ties the environment's quantum oracle interpretation to $U$, requires $0 < t$, and requires the local quantum input to be the promised eigenstate; \lean{initialization} admits every initial state.
The returned $t$ bits encode an integer $k$ in big-endian order, with the most significant bit first, and the decoded phase estimate is $\widehat\theta = k/2^t$.
The relation \lean{accepts} requires $d_{\mathbb R/\mathbb Z}(\widehat\theta,\theta) \le 2^{-t}$, where the distance is circular and phases are interpreted modulo one.
The threshold \lean{successLB} is $4/\pi^2$, and \lean{states} is empty, because what the algorithm must leave behind is the borrowed eigenstate, and that is the separate restoration obligation described in Section~\ref{sec:borrowed}.
The execution model offers Toffoli, CNOT and arbitrary single-qubit gates as native gates, queries to $U$ and to controlled-$U$, and operations on bits and words.

The event budgets are given by explicit formulas: $2^t - 1$ controlled-$U$ queries, $2t + 3\binom{t}{2}$ single-qubit gates, $2\binom{t}{2} + 3\lfloor t/2 \rfloor$ CNOT gates, $t$ qubit preparations, $t$ single-qubit $Z$ measurements and one bit-packing operation.
The gate counts are those of a Hadamard gate on each of the $t$ phase qubits followed by the inverse quantum Fourier transform written in the native alphabet: each of its $\binom{t}{2}$ controlled phases costs three single-qubit gates and two CNOT gates, and each of its $\lfloor t/2 \rfloor$ swaps costs the three CNOT gates of Figure~\ref{fig:swap}.
The theorem \lean{code\_events} in \leanpath{QuantumAlg/Algorithms/QPE/Resources.lean} separately proves that every path of the exhibited implementation attains these event counts; the general \lean{Solves} contract requires only that counts stay within the budgets.

Resident space is $t + n$ qubits and $2t$ classical bits, and auxiliary space is $t$ qubits and $2t$ bits.
At positive width, the specification sets the quantum-time limit to $2^t + 2t + 5\binom{t}{2} + 3\mathbf{1}_{t \ge 2} + 2$, where the indicator is one when at least one swap is needed.
The disjoint swaps run in parallel and contribute three layers together; charging those swaps serially gives the looser upper bound $2^t + 2t + 5\binom{t}{2} + 3\lfloor t/2 \rfloor + 2$.
Both bounds are conservative for quantum time: they come from an analysis that charges every event one unit, whereas the canonical quantum duration charges only native quantum gates and quantum queries.
Each of these numbers is an explicit formula in the specification module, stated independently of the circuit.

\subsection{\lean{Algorithm}: one run}
\label{sec:algorithm}

Listing~\ref{lst:algorithm-run} defines an \lean{Algorithm} to contain exactly one \lean{AlgorithmRun}, which pairs a circuit with a declaration of every live final resource and of the storage from which its classical result is read.
The run is the sole source of semantics, outputs, paths and resource use.
The execution distribution, the output distribution, the relational distribution over initial value and result, the event multiset of each path and the path probabilities are all derived from \lean{run} by the framework, and none of them can be supplied independently of the run.

\begin{leanlisting}{\lean{AlgorithmRun} and \lean{Algorithm}}{lst:algorithm-run}
-- source: QuantumAlg/Core/Algorithm/AlgorithmRun.lean:38-44
public structure AlgorithmRun (model : ExecutionModel) (pre post : RuntimeCtx model.kind)
    (outputShape : CShape) (allowAllocation : Bool := true) : Type 1 where
  code : Circuit model pre post allowAllocation
  output : Output post outputShape

-- source: QuantumAlg/Core/Algorithm/Problem.lean:83-87
public structure Algorithm (model : ExecutionModel) (pre post : RuntimeCtx model.kind)
    (outputShape : CShape) (allowAllocation : Bool := true) : Type 1 where
  run : AlgorithmRun model pre post outputShape allowAllocation
\end{leanlisting}

The declaration of the final storage makes the observation explicit.
The declared keys must be a permutation of the live storage keys of the output context.
An algorithm that omits a live qubit or classical cell from its output, repeats a key or invents one cannot be formed; its classical result is read from that live storage.

\subsection{\lean{Solves}: seven obligations on the same execution}
\label{sec:solves}

The relation \lean{Solves} in Listing~\ref{lst:solves} is a proposition about an algorithm and a problem.
Its seven fields are quantified identically: over every environment, every value of the problem's \lean{Witness} type, every external reference register and every normalized input whose local state satisfies \lean{valid} and whose joint initialization satisfies \lean{initialization}.
The candidate remains fixed across all such admitted combinations.

\begin{leanlisting}{The \lean{Solves} relation}{lst:solves}
-- source: QuantumAlg/Core/Algorithm/Solves.lean:35-92
public structure Solves {allowAllocation : Bool} {model : ExecutionModel}
    {pre post : RuntimeCtx model.kind} {outputShape : CShape}
    (algorithm : Algorithm model pre post outputShape allowAllocation)
    (problem : Problem model pre post outputShape) : Prop where
  success : ∀ (environment : model.Environment) (witness : problem.Witness) (Reference : Register)
    (input : Execution.OracleInput model.classicalOracles pre Reference),
    problem.valid environment witness input.caller.localState →
    problem.initialization environment witness input.localState →
      (problem.successLB environment witness).value ≤
      (algorithm.relationalDistribution environment input).toOuterMeasure
        {pair | problem.accepts environment witness pair.1 pair.2}
  states : ∀ (environment : model.Environment) (witness : problem.Witness) (Reference : Register)
    (input : Execution.OracleInput model.classicalOracles pre Reference),
    problem.valid environment witness input.caller.localState →
    problem.initialization environment witness input.localState →
      ∀ requirement ∈ problem.states environment witness,
      requirement.Holds input.caller (algorithm.run.finalState environment Reference input.state)
  borrowed : ∀ (environment : model.Environment) (witness : problem.Witness) (Reference : Register)
    (input : Execution.OracleInput model.classicalOracles pre Reference),
    problem.valid environment witness input.caller.localState →
    problem.initialization environment witness input.localState →
      algorithm.run.RestoresBorrowed environment input
  events : ∀ (environment : model.Environment) (witness : problem.Witness) (Reference : Register)
    (input : Execution.OracleInput model.classicalOracles pre Reference),
    problem.valid environment witness input.caller.localState →
    problem.initialization environment witness input.localState →
      ∀ path,
      algorithm.run.Supports environment Reference input.state path →
        algorithm.run.code.resourcePool path ≤ problem.resources.events
  resident : ∀ (environment : model.Environment) (witness : problem.Witness) (Reference : Register)
    (input : Execution.OracleInput model.classicalOracles pre Reference),
    problem.valid environment witness input.caller.localState →
    problem.initialization environment witness input.localState →
      ∀ path,
      algorithm.run.Supports environment Reference input.state path →
        algorithm.run.code.peakResidentSpace path ≤ problem.resources.resident
  auxiliary : ∀ (environment : model.Environment) (witness : problem.Witness) (Reference : Register)
    (input : Execution.OracleInput model.classicalOracles pre Reference),
    problem.valid environment witness input.caller.localState →
    problem.initialization environment witness input.localState →
      ∀ path,
      algorithm.run.Supports environment Reference input.state path →
        algorithm.run.code.peakAuxiliarySpace path ≤ problem.resources.auxiliary
  time : ∀ (environment : model.Environment) (witness : problem.Witness) (Reference : Register)
    (input : Execution.OracleInput model.classicalOracles pre Reference),
    problem.valid environment witness input.caller.localState →
    problem.initialization environment witness input.localState →
      ∀ path,
      algorithm.run.Supports environment Reference input.state path →
        algorithm.quantumTime path ≤ problem.quantumTimeLimit
\end{leanlisting}

The obligations fall into three groups.
\lean{success} and \lean{states} are the correctness obligations: the relational event has probability at least the threshold, and every listed state requirement holds for the actual final state.
\lean{borrowed} is the restoration obligation described in Section~\ref{sec:borrowed}.
The remaining fields, \lean{events}, \lean{resident}, \lean{auxiliary} and \lean{time}, are the resource obligations.
They range over the paths of the run's circuit that the run \lean{Supports}, that is, the paths of positive probability, and compare the event multiset, the space peaks and the quantum time computed for each path with the budgets of the problem.

The property that carries the paper is visible in the listing: every one of the seven fields is a statement about the single run \lean{algorithm.run}.
The success field reads the relational distribution derived from it, the state and restoration fields read its final state, and the four resource fields read the paths of its circuit \lean{algorithm.run.code}.
A resource bound in a \lean{Solves} certificate is therefore a bound on the circuit whose success probability was proved, on the same inputs, and no field admits a separate account of the run.

\subsection{The final theorem}
\label{sec:main-theorem}

Each executable algorithm exports a final theorem \lean{main} stating that some algorithm solves its problem.
Quantum phase estimation gives a representative instance in Listing~\ref{lst:qpe-main}.
For each $n$ and $t$, the same candidate is certified for all admitted QPE instances.
Its proof, omitted from the listing, supplies the exported \lean{algorithm} as the candidate together with a theorem that this algorithm solves the problem, and the proof of that theorem discharges the seven obligations in the order of Listing~\ref{lst:solves}.

\begin{leanlisting}{The final theorem for quantum phase estimation (proof omitted)}{lst:qpe-main}
-- source: QuantumAlg/Algorithms/QPE/Endpoint.lean:141-145
public theorem main (n t : Nat) :
    ∃ candidate : Algorithm (model n) (pre t n) (post t n) (.bits t),
      Algorithm.Solves candidate (problem n t) :=
\end{leanlisting}

Although \lean{main} quantifies over every width $t$, its algorithm guarantee applies to admitted inputs, which require $t > 0$.
At $t = 0$, the predicate \lean{valid} admits no inputs, so the theorem provides no nontrivial algorithm guarantee at that width.

The same form applies to order finding, discrete logarithms and the quantum attacks on RSA-2048 and P-256.
Separate theorems may give stronger facts about the implementation exhibited by the proof, such as exact gate counts or closed-form space bounds.
The common \lean{main} statement nevertheless has one role: it asserts the existence of an implementation satisfying every obligation of the stated problem.

The final theorem certifies existence, and its proof identifies an implementation that meets the contract.
Optimality, uniqueness and gate-for-gate agreement with a textbook circuit lie outside the statement, and so does any property of the execution model beyond what the model's own definitions state.
To know what has been certified, a reviewer reads the statement of \lean{main}, the \lean{Problem} it names and the execution model; Section~\ref{sec:review} makes this precise.
\section{Execution and resource accounting}
\label{sec:execution}

Every obligation in Listing~\ref{lst:solves} is evaluated on objects that the framework computes from the circuit and its execution: distributions, final states, event multisets, space peaks and quantum time.
This section describes where those objects come from, because the trust a reviewer places in a certificate is exactly the trust placed in these definitions.

\subsection{Model-owned tokens and the trust boundary}
\label{sec:tokens}

An execution model fixes four alphabets: the native gates, the quantum oracles, the classical operations and the classical oracles.
A token is an element of one of these alphabets, and each alphabet is a type owned by its declaration: a gate token carries its operator and its event coordinate, a quantum oracle token carries its event coordinate and receives its operator from the environment, and a classical operation carries its semantics on the classical store and its coordinate.
No constructor turns an arbitrary unitary and a label into a gate.

The atomic unitary operations of a circuit are the leaves of Listing~\ref{lst:leaf}.
A leaf is one token of the model applied to its actual wires, and its resource event is fixed by that token.
Its operator is the gate operator or, for an oracle query, the operator selected from the environment's interpretation of that declared oracle.
Within a fixed execution model, relabeling a multi-qubit subroutine as a native gate, or charging a query to an oracle that the model does not declare, would require a term that cannot be formed, so the boundary between what is native and what must be built is enforced by types.

\begin{leanlisting}{Atomic unitary operations, their events and their operators}{lst:leaf}
-- source: QuantumAlg/Core/Base/UnitaryCircuit/Leaf.lean:26-32
public inductive UnitaryLeaf (gates : GateModel) (oracles : OracleModel)
    {count : Nat} (layout : QuantumLayout count) where
  | gate (application : GateModel.Application gates layout)
  | oracle (application : OracleModel.Application oracles layout)

-- source: QuantumAlg/Core/Base/UnitaryCircuit/Leaf.lean:39-55
@[expose]
public def event (leaf : UnitaryLeaf gates oracles layout)
    (operations : ClassicalOpModel := .empty)
    (classicalOracles : ClassicalOracleModel := .empty) :
    Resource.Event (.quantum gates oracles operations classicalOracles) :=
  match leaf with
  | .gate application => .gate application.coordinate
  | .oracle application => .oracle application.coordinate

@[expose]
public noncomputable def operator (environment : oracles.Interpretation) :
    UnitaryLeaf gates oracles layout →
    UnitaryOperator layout.register
  | .gate application => application.operator
  | .oracle application => application.operator environment
\end{leanlisting}

\begin{figure}[t]
  \centering
  \includegraphics[width=0.8\linewidth]{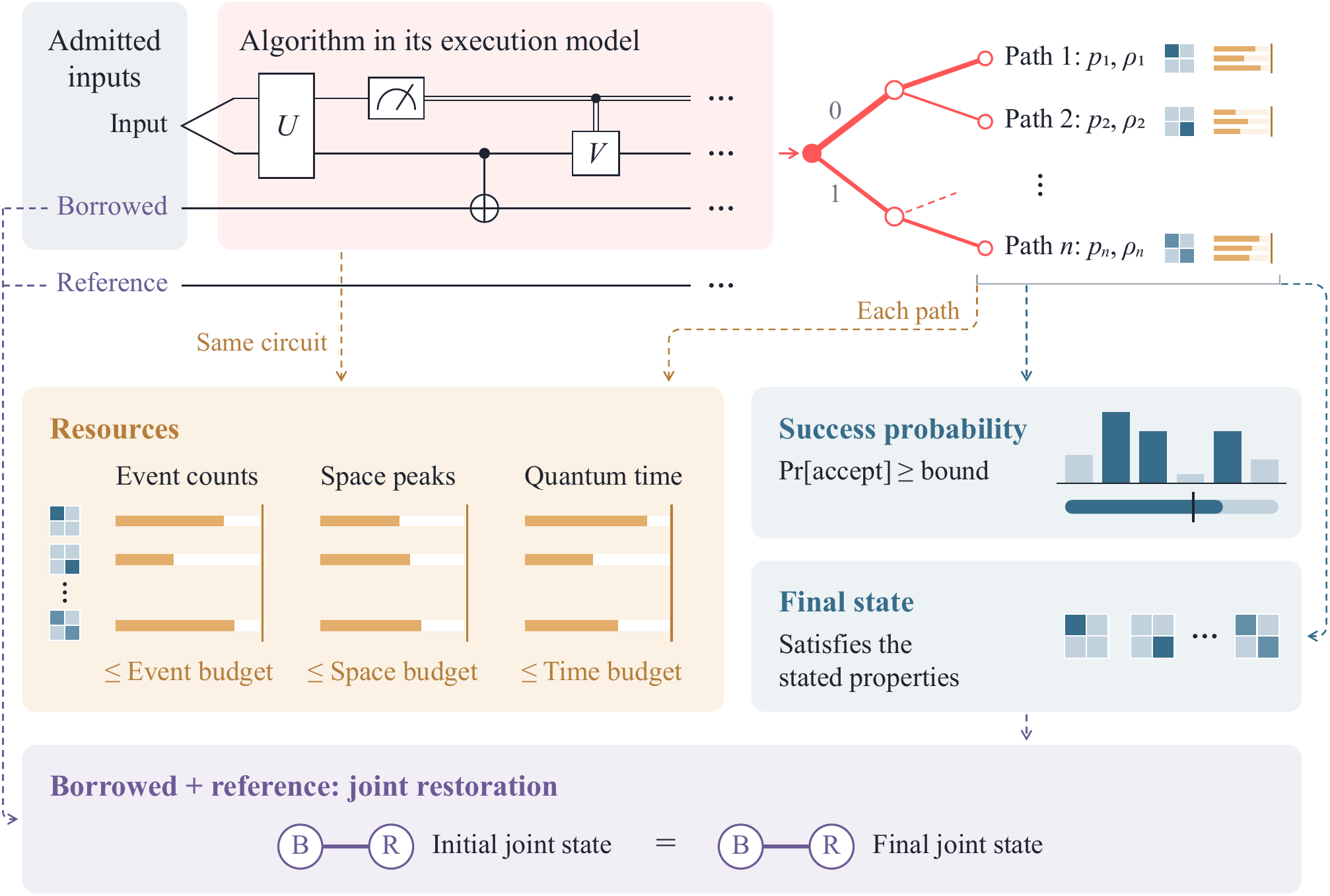}
  \caption{One execution supplies the guarantees in \lean{Solves}.
    The schematic circuit acts on the admitted inputs, leaves the external reference $R$ untouched and branches on measurement outcomes into paths with probabilities $p_i$ and final states $\rho_i$.
    Thicker lines mark more probable paths.
    Combined over the paths, the probability of the accepted results, drawn dark, must reach the success threshold, and the final state must satisfy the properties stated in \lean{Problem}.
    Dashed arrows show dependencies, including the two inputs to resource accounting: the same circuit and each path, with the event, space and time budgets applying to every path of positive probability.
    At the bottom, equality of the initial and final joint states of borrowed systems $B$ and the reference $R$ expresses the restoration condition fixed by the framework.}
  \label{fig:execution}
\end{figure}

\subsection{One interpreter, every observation derived}
\label{sec:interpreter}

Circuits combine unitary leaves with qubit preparations, measurements, classical operations, classical queries and control that depends on measured values.
One interpreter executes a circuit under an environment that supplies its oracle interpretations, on a classical--quantum state whose quantum register is extended by an arbitrary external reference.
The result is a state indexed by the final classical store, the hidden service state, the initial store and the path taken.
Final states and probabilities are derived from this result, while resource use is computed from the same circuit and its paths.
Figure~\ref{fig:execution} separates these dependencies and shows how the initial and final joint states enter the restoration condition.

Probabilities are traces of blocks of that state, so the execution distribution, the output distribution and the relational distribution are projections of one normalized object.
Resource obligations are stated for every execution path with positive probability.
A branch that measurement never reaches does not bind the budget, and a branch that is reached with any positive probability does.

\subsection{Events, space and quantum time}
\label{sec:accounting}

The events of a path form a multiset over the coordinates of the model: gate applications, oracle queries, preparations, measurements, classical operations and classical queries.
A budget bounds each coordinate separately, so an algorithm cannot trade a controlled-$U$ query for a single-qubit gate, and the classical steps that a paper usually relegates to prose are events in the same multiset as the quantum gates.

Where event counts accumulate operations, space records the peaks of live storage along the path.
The \emph{resident space} of a path is the peak number of live classical bits and of live quantum systems of each dimension, and its \emph{auxiliary space} is the peak of the live storage whose declared role is auxiliary.
Borrowed systems, including the eigenstate used in phase estimation, count toward resident space but not owned auxiliary space.

Quantum computation time uses the dependency structure of the same complete execution path.
The canonical duration \lean{Resource.Event.quantumDuration} assigns one unit to each native quantum gate and quantum-oracle query, and zero to measurements, fresh zero preparations, classical operations and classical queries.
Preparation or reset routines that apply quantum gates still incur the time of those gates.
Thus, zero duration leaves an operation in the event count while excluding it from this particular time measure.

Sequential execution adds quantum times, and iteration includes the time of every recorded repetition.
Explicit parallel composition of disjoint circuits takes the maximum of their times; a measurement-dependent branch contributes the time of the selected continuation.
The framework derives \lean{Algorithm.quantumTime} from the circuit and path, and \lean{Solves.time} compares it with \lean{Problem.quantumTimeLimit} on every execution path with positive probability.
General event-weighted depth remains available through \lean{Circuit.depth}, but the algorithm contract fixes the quantum duration and allows the problem to choose only its upper bound.
Consequently, the time bound depends on which gates and queries the execution model admits, and carries no claim about elapsed time on hardware.

\subsection{Borrowed systems are restored jointly}
\label{sec:borrowed}

A borrowed system is one that the algorithm receives, uses and must hand back.
In \LeanQuantumAlg{} its restoration is an obligation of its own: the relation \lean{RestoresBorrowed} in Listing~\ref{lst:borrowed} compares the complete borrowed classical--quantum--reference state after execution with the one before, as a single joint object.
\lean{Solves} requires this restoration for every admitted input, independently of the user-specified list of state requirements.
This equality concerns the final joint state obtained by combining the measurement outcomes; restoration conditional on an individual outcome would be a separate claim.
The resource obligations, in turn, explicitly bound every execution path with positive probability.

\begin{leanlisting}{Borrowed restoration and paths with positive probability}{lst:borrowed}
-- source: QuantumAlg/Core/Execution/Borrowed/State.lean:171-179
public def BorrowedRestored {model : ExecutionModel} {pre post : RuntimeCtx model.kind}
    {allowAllocation : Bool} (code : Circuit model pre post allowAllocation) (Reference : Register)
    {H K : Type} [Fintype H] [Fintype K]
    (initial : ExecutionState (Execution.Store pre × H) (Register.prod pre.register Reference))
    (final : ExecutionState (Execution.Store post × K) (Register.prod post.register Reference)) :
    Prop :=
  Execution.observeBorrowedAfter code Reference final = Execution.observeBorrowed Reference initial

-- source: QuantumAlg/Core/Algorithm/AlgorithmRun.lean:170-183
public def RestoresBorrowed (run : AlgorithmRun model pre post outputShape allowAllocation)
    (environment : model.Environment)
    {Reference : Register}
    (input : Execution.OracleInput model.classicalOracles pre Reference) : Prop :=
  run.code.BorrowedRestored Reference input.state (run.execute environment Reference input.state)

public def Supports (run : AlgorithmRun model pre post outputShape allowAllocation)
    (environment : model.Environment)
    (Reference : Register)
    (input : ExecutionState (Execution.Store pre × model.HiddenState)
      (Register.prod pre.register Reference))
    (path : run.code.Path) : Prop := 0 < run.pathMass environment Reference input path
\end{leanlisting}

The joint comparison is what makes the obligation meaningful.
If the borrowed qubit is half of a Bell pair whose other half is the external reference, an algorithm that returns the qubit in the correct marginal state but decorrelated from the reference satisfies every requirement on the marginal and fails \lean{RestoresBorrowed}.
Phase estimation on a borrowed eigenstate preserves its joint state with the reference and therefore satisfies the restoration obligation.
\section{Review under the framework}
\label{sec:review}
\label{sec:reviewer-reads}

A checked proof establishes the formal statement presented to Lean.
Accepting an agent-written development also requires a human to establish that this statement expresses the intended claim.
The framework concentrates this second task in the final theorem, its specification, its execution model and the shared definitions that connect them.
The fraction of code requiring human review in Section~\ref{sec:overview} measures the source lines involved, depending on which shared definitions have already been reviewed; it does not measure the difficulty of understanding those lines.

Review begins with \lean{main}: its parameters and hypotheses determine the scope of the existence claim, and the named \lean{Problem} determines what an implementation must satisfy.
The reviewer then compares that problem and the definitions used to construct it with the source claim, checking its \lean{Witness} type, how \lean{valid} relates witnesses to admitted environments, the admitted inputs and initialization, the acceptance relation and success threshold, the final-state requirements and the resource limits.
In particular, a narrower input promise, a weaker acceptance condition or a larger resource budget can make a theorem easier to prove while changing its meaning.
The interface permits independently stated requirements, but their independence from a particular candidate is established by reviewing these definitions.
If a budget is defined from a candidate's own cost, \lean{Solves} still checks the actual execution against that budget; human review determines whether this is the intended resource claim.

The execution model supplies the next part of that meaning.
Its gate alphabet, oracle declarations and classical operations determine which actions the algorithm can take as primitives, while the environment supplies the interpretations of the declared oracles.
Consequently, the reviewer checks that the granted operations and their semantics match the assumptions of the source claim.
The resource conventions of Section~\ref{sec:accounting} then determine how those operations contribute to event counts, space and quantum time.
A bound in one model need not describe the cost of realizing its primitives in another model or on hardware.

Finally, the shared definitions of Section~\ref{sec:framework} and Section~\ref{sec:execution} must be trusted or reviewed.
They determine how an execution produces its observations and how \lean{Solves} compares them with the problem.
Once this correspondence has been established, Lean checks that the implementation supplied by the proof satisfies all seven obligations on the same admitted inputs.
The reviewer can then leave the candidate construction and its proof to Lean; definitions used to state the problem remain part of the review.
Claims about a particular textbook construction, optimality or a physical realization require their own statements and evidence.
Appendix~\ref{appendix:drift} illustrates the division between enforced consistency and human judgment through concrete ways in which a formal claim can depart from its source.

This review also presupposes the usual validation of a Lean development.
A consumer checks whether the exported theorem depends on \lean{sorryAx} and audits its remaining axiom dependencies, following the checks described in the Lean reference~\cite{lean2026validating}.
These standard checks establish the proof's dependency assumptions; they do not compare the formal specification with its informal source.
\section{Conclusion}
\label{sec:conclusion}

In this work, we present \LeanQuantumAlg{}, which provides a common interface for certifying quantum algorithms against independently stated claims, by introducing a \lean{Problem} that records the commitments of a claim, an \lean{Algorithm} that supplies one execution, and \lean{Solves} that relates the two.
To our knowledge, \LeanQuantumAlg{} is the first framework in which one checked theorem binds correctness, borrowed-state restoration and the quantum and classical resource bounds of a complete execution, including its repetitions and classical processing.
The nine algorithms show how this interface accommodates both textbook examples and quantum attacks on cryptographic problems, with event, space and quantum-time bounds interpreted in an explicit execution model.
Once the shared framework has been audited, the code requiring human review accounts for less than 1\% of the lines of code in the import closure of each algorithm.
This organization gives human review a defined subject: whether the theorem, specification and model express the intended claim, while Lean checks that the supplied implementation satisfies it.

Two aspects of the current development call for further work.
First, the common existential form of \lean{main} identifies the problem being solved, but understanding its mathematical content still requires following Lean definitions, parameters and type-level representations.
Clearer names, simpler statement forms and explanations tied to the checked declarations would expose the input assumptions, acceptance criterion, state requirements and resource formulas more directly.
Second, parts of the AI-built development use indirect constructions, redundant intermediate definitions or repeated arguments.
Kernel checking establishes the resulting theorems under their declared assumptions, but it does not make these constructions easy to maintain; refactoring shared arguments, removing unnecessary layers and shortening proof dependencies would make the library easier to extend while preserving its statements and the execution semantics on which their checked guarantees depend.

More broadly, \LeanQuantumAlg{} moves the formal verification of quantum algorithms from handcrafted, circuit-level verification toward an industrial-scale paradigm in which AI agents generate certified algorithms and humans perform a lightweight audit of their statements.
When AI agents propose new quantum algorithms, the same contract lets Lean check each proposal against a problem that humans have reviewed, opening a path toward AI-driven discovery of quantum algorithms.

\textbf{Acknowledgment.}---This work was partially supported by the National Natural Science Foundation of China (Grant. No.~92576114, 12447107), the Guangdong Provincial Quantum Science Strategic Initiative (Grant No.~GDZX2403008, GDZX2503001).

\appendix
\section{Evidence for Table~\ref{tab:comparison}}
\label{appendix:evidence}

The rows of Table~\ref{tab:comparison} are formal tools described in public papers that identify the versions compared; each row is judged against the papers cited in the table and, where a paper defers to it, the tool's official documentation.
In the table, \cmark{} means that the compared version provides the capability, \pmark{} means that it provides a restricted form, and \xmark{} means that the cited version does not show the capability; QWIRE's authors, for example, list a gate counter only as an operation expressible in the host language~\cite{paykin2017qwire}.
The Expected output and Success probability columns require proofs of functional correctness and of a success-probability bound, respectively.
The Dynamic circuit column requires mid-circuit measurement with classical feedback inside the certified program, and the Same object column requires the correctness and resource bounds to be proved for one object.
For every resource column, a bound must be established by the work's formal machinery to qualify.
The Gates column requires a certified count of primitive gates, and the Qubits column requires a bound on the qubits used, including auxiliary qubits.
The Depth/Time column requires either a bound on circuit depth under the cited work's metric or a bound on the quantum computation time of every complete execution path.
Apart from this work, none of the compared tools bounds quantum computation time, so their marks in this column follow from their depth results.
The Classical bits column requires a bound on classical memory, and the Classical operations column requires a bound on classical computation.
A type-system soundness result and a proof checked in a proof assistant have different trust bases; the marks do not rank them.

\paragraph{Qbricks~\cite{chareton2021automated}.}
Qbricks is an automated deductive verification framework embedded in Why3: Qbricks-DSL builds parametric unitary circuits, and Qbricks-Spec with the HQHL logic specifies and discharges functional, probabilistic and size obligations; the verified algorithms include Shor's order finding, phase estimation and Grover search.
Dynamic circuit is \xmark{} because measurement is stated to be out of the scope of the language; probabilities are reasoned about through a terminal-measurement predicate.
Same object is \cmark{} because functional and \texttt{size} specifications attach to the same Qbricks-DSL term.
Gates is \cmark{}: the \texttt{size} operator, whose equational theory is part of the logic, counts primitive gates, and an oracle is a concrete circuit~$C$ whose gates enter that count, as in the Deutsch--Jozsa bound $2n+\texttt{size}(C)+2$.
Qubits is \cmark{}: \texttt{width} counts the input and output wires, and \texttt{ancillas} counts the auxiliary qubits.
Depth/Time is \xmark{} because the cited work establishes size and width bounds, with no depth or time operator.
Classical bits is \xmark{} because \texttt{width} counts quantum wires and the ancilla construct creates qubits.
Classical operations is \xmark{} because the cited resource bounds concern the generated quantum circuits; they do not bound the cost of circuit construction or measurement-dependent classical post-processing.

\paragraph{SQIR.}
SQIR~\cite{hietala2021verified} is a Coq-embedded circuit language with a verified optimizer, VOQC.
Ref.~\cite{hietala2021proving} proves algorithm correctness including probabilistic bounds, and Ref.~\cite{peng2023formally} certifies an end-to-end hybrid implementation of Shor's algorithm through SQIR and the classical language RCIR.
The row covers the three publications together.
Dynamic circuit is \cmark{} through full SQIR's measurement construct, which branches on a mid-circuit outcome~\cite{hietala2021proving}.
Same object is \cmark{} because the same Shor circuit object has a correctness proof, a success bound and a gate-count bound~\cite{peng2023formally}.
Gates is \cmark{}: the certified resource statement bounds the primitive gate count (\texttt{ugcount}) of the Shor circuit, whose modular-multiplication oracle is a concrete circuit~\cite{peng2023formally}.
Qubits is \cmark{}: the number of qubits is bounded by the maximal index of the global register, including the ancillae of the modular arithmetic~\cite{peng2023formally}.
Depth/Time is \xmark{} because depth appears only as a metric of the optimizer, without a certified bound~\cite{hietala2021verified}.
Classical bits is \xmark{} because RCIR programs on classical bits are translated into SQIR qubits before resources are counted, so the certified resources are qubits and gates only.
Classical operations is \xmark{}: the classical post-processing is proved correct and extracted to OCaml, but its cost is not a resource metric; the only iteration bound is a correctness lemma for the continued-fraction step.

\paragraph{QWIRE.}
QWIRE~\cite{paykin2017qwire} is a linearly typed core circuit language with a host-language interface; its denotational superoperator semantics is formalized in Coq~\cite{rand2018qwire}.
Dynamic circuit is \cmark{} through dynamic lifting, where the host chooses the continuation after a measurement~\cite{paykin2017qwire}.
Same object is \xmark{} because no resource bound is proved of a boxed circuit, and Gates is \xmark{} for the same reason.
Qubits is \pmark{}: linear wire types track the bits and qubits of an interface but may hide internal ancillae, and no width is computed~\cite{rand2018qwire}.
Classical bits is \xmark{}: \texttt{Bit} is an interface wire type interpreted as a dimension, not a memory resource that is bounded.
Depth/Time and Classical operations are \xmark{} because no certified depth or classical-operation bound is given; a gate counter is listed as a possible host-language extension only~\cite{paykin2017qwire}.

\paragraph{CoqQ~\cite{zhou2023coqq}.}
CoqQ is a foundational verifier in Coq for the imperative language qwhile, with Hilbert-space semantics, labeled Dirac notation and a verified quantum Hoare logic; the algorithms it verifies include Grover search, phase estimation, HHL and an algorithm for the hidden subgroup problem.
Dynamic circuit is \cmark{} because conditionals and loops branch on measurement outcomes.
Same object is \xmark{} because these algorithms are verified through Hoare triples only, without resource bounds, which also settles Gates, Depth/Time and Classical operations as \xmark{}.
Qubits is \pmark{}: variables of arbitrary finite types fix the Hilbert-space dimension, but there is no width or ancilla analysis.
Classical bits is \xmark{} because qwhile has classical control but quantum data only; classical variables are named as future work.

\paragraph{QEDIA~\cite{ren2026qedia}.}
Expected output and Success probability are \cmark{}: QEDIA formalizes a quantum Hoare logic for programs with classical variables in Lean and proves its proof system sound for partial and total correctness.
Its repeat-until-success example has success probability at least $1-4^{-N}$, where $N$ is the maximum number of attempts.
Dynamic circuit is \cmark{} because a measurement result controls the subsequent loop in that example.
Same object is \xmark{} because these correctness results are not accompanied by a resource bound for the same program, and Gates is \xmark{} because no gate-count metric is established.
Qubits is \pmark{}: each quantum variable has a finite, nonempty set of basis labels, so its type fixes a Hilbert-space dimension, including that of the ancilla in the repeat-until-success example, but no peak-width or auxiliary-space bound is established.
Classical bits is \xmark{} because classical variables and feedback are represented semantically, without a certified bound on classical memory.
Depth/Time and Classical operations are \xmark{} because the cited version establishes no certified depth or classical-operation metric.
The termination proof and the loop parameter do not supply these resource guarantees.

\paragraph{QuRA.}
QuRA~\cite{colledan2025circuit} provides type-based static width analysis for Quipper-style circuit-description programs.
Its extension based on Proto-Quipper-RA uses effect and refinement types to derive parametric upper bounds on width, depth and gate count, and the tool infers bounds for global metrics automatically~\cite{colledan2025flexible}.
The following cells refer to this extended analysis.
Expected output, Success probability and Same object are \xmark{} because the correctness result is soundness of the resource types, not functional correctness of the analyzed program.
Dynamic circuit is \pmark{} because measurement is an elementary circuit operation from qubits to bits, while Quipper-style dynamic lifting is left to other variants of the language.
Gates is \cmark{} because gate count is a global metric of the type system over an arbitrary but fixed set of elementary operations~\cite{colledan2025flexible}.
Qubits is \cmark{} because the width, the peak number of live wires including internal ancillae, bounds the live qubits~\cite{colledan2025flexible}.
Depth/Time is \cmark{} because Ref.~\cite{colledan2025flexible} proves a soundness theorem for depth as a local metric; the same publication reports that the tool's implementation of local metrics was still in progress.
This depth analysis already includes repetitions within the generated circuit.
Classical bits is \cmark{} because the wires counted by the width include bits, and the implementation also provides a dedicated number-of-bits metric.
Classical operations is \pmark{} because the gate count includes measurements and operations on classical wires, while the cost of host-language circuit construction and of classical work outside the circuit is out of scope.

\paragraph{\LeanQuantumAlg{}.}
The row for this work follows from Section~\ref{sec:framework} and Section~\ref{sec:execution}: correctness and success-probability obligations, mid-circuit measurement with control on measured values, and a single run for all obligations give the first four marks.
Gates, Qubits, Classical bits and Classical operations are \cmark{} through event budgets and resident and auxiliary space bounds over quantum systems and classical bits.
A budget bounds gate applications and oracle queries as separate event coordinates, and resident space counts the live quantum systems of each dimension, so qubits and qudits are bounded alike.
Depth/Time is \cmark{} through quantum time, which we regard as the more suitable measure for a complete execution.
Quantum time, defined in Section~\ref{sec:accounting}, follows the dependencies of each execution path, including every repetition and the continuation selected after a measurement, and \lean{Solves.time} bounds it on every path with positive probability.
General event-weighted depth is available separately at the \lean{Circuit} level.
The framework attaches no fault-tolerant or physical resource estimate to a certificate.
\section{Measuring the fraction of code requiring human review}
\label{appendix:footprint}

Table~\ref{tab:footprint} measures the nine algorithms using the same library version as the listings in Section~\ref{sec:framework}.
Every measured final theorem has the common existential \lean{main} form.

Two scripts in the paper repository produced Table~\ref{tab:footprint}: \texttt{StatementClosure.lean} and \texttt{loc\_footprint.py}.
The Lean program is run with \texttt{lake env lean --run} on the compiled \LeanQuantumAlg{} library, and the Python driver calls it once per algorithm, counts source lines and writes the table.
They write the Lean requests and responses, per-module and per-endpoint CSV files, a JSON summary and the table's TeX body.

Measurement starts from the module that exports an algorithm's final theorem \lean{main}.
The nine modules and theorems are:
\begin{itemize}[nosep]
  \item Deutsch--Jozsa: \leanpath{QuantumAlg.Algorithms.DeutschJozsa.Endpoint},\\
    theorem \lean{QuantumAlg.DeutschJozsa.main};
  \item Bernstein--Vazirani: \leanpath{QuantumAlg.Algorithms.BernsteinVazirani.Endpoint},\\
    theorem \lean{QuantumAlg.BernsteinVazirani.main};
  \item Grover search: \leanpath{QuantumAlg.Algorithms.Grover.Endpoint},\\
    theorem \lean{QuantumAlg.GroverSearch.main};
  \item quantum phase estimation: \leanpath{QuantumAlg.Algorithms.QPE.Endpoint},\\
    theorem \lean{QuantumAlg.QuantumPhaseEstimation.main};
  \item amplitude estimation: \leanpath{QuantumAlg.Algorithms.AmplitudeEstimation.Endpoint},\\
    theorem \lean{QuantumAlg.AmplitudeEstimation.main};
  \item order finding:\\
    \leanpath{QuantumAlg.Algorithms.OrderFinding.General.NativeEndpoint.Canonical},\\
    theorem \lean{QuantumAlg.OrderFinding.ShorSourceJoint.SourceNativeEndpoint.Canonical.main};
  \item finite-cyclic discrete logarithm:
    \leanpath{QuantumAlg.Algorithms.FiniteCyclicDLP.Binary.Endpoint},\\
    theorem \lean{QuantumAlg.FiniteCyclicDLP.BinaryEndpoint.main};
  \item quantum attack on RSA-2048: \leanpath{QuantumAlg.Algorithms.Factoring.RSA2048},\\
    theorem \lean{QuantumAlg.Factoring.RSA2048.main};
  \item quantum attack on P-256: \leanpath{QuantumAlg.Algorithms.EllipticCurve.P256},\\
    theorem \lean{QuantumAlg.EllipticCurve.P256LogicalResources.main}.
\end{itemize}

Each row of Table~\ref{tab:footprint} cites first the paper in which the algorithm originates and then, where it differs, the source against which the Lean statement is registered: the lecture notes~\cite{dewolf2019lecturenotes} for Deutsch--Jozsa, Bernstein--Vazirani and Grover search, the lecture notes~\cite{lin2022qasc} for phase estimation and amplitude estimation, Ref.~\cite{shor1997polynomial} for order finding and, together with the uniform-superposition preparation of Ref.~\cite{shukla2024uniformsuperposition}, for the finite-cyclic discrete logarithm, the short-discrete-logarithm route~\cite{ekera2017quantum} and its success-probability analysis~\cite{ekera2023shortdlp} for RSA-2048, and the elliptic-curve algorithm~\cite{proos2004shor} with the resource analysis of Ref.~\cite{roetteler2017quantum} and the P-256 domain parameters~\cite{nist2023sp800186} for P-256.
Before collecting new Lean responses, the driver checks with \texttt{lake --no-build build} that every module exporting \lean{main} and its import closure are compiled and up to date, so that the constants inspected are those of the sources counted.
The driver can also be pinned to a revision of the library: it then refuses to run unless the checkout is at that revision without local changes to the library sources, and it records the revision in every Lean request, so that saved responses are never reused against different sources.

\paragraph{Statement dependencies.}
The Lean program imports the entire import closure of the module exporting \lean{main}, including private declarations, and starts from the constants that occur in the type of \lean{main}.
It then follows, for every constant reached, the constants in its type and, for definitions, instances, structures, inductive types (with their constructors and projections), recursors and opaque constants, the constants in its body.
Two kinds of constants are leaves whose bodies are never followed.
Theorems and axioms are recorded and never expanded, so a theorem reached as a proof term inside a definition contributes nothing beyond its presence, and the proof of \lean{main} is never visited.
A constant whose type concludes with \lean{Algorithm} contributes only the constants of its type: a reviewer reads its signature, and Lean checks its body through \lean{Solves}.
Constants defined outside the \lean{QuantumAlg} library, in Mathlib or in the Lean core, are recorded and not expanded either, since they cannot mention the library.
The result is the set of library declarations that fix the meaning of \lean{main}.
The program also checks that each measured statement of \lean{main} has the existential form of Section~\ref{sec:main-theorem}.

The signature-only treatment of an \lean{Algorithm} is appropriate for human review when the problem can be understood independently of its implementation.
If a budget or acceptance condition is constructed from an algorithm, the reviewer also follows the definitions needed to interpret that dependency.
The \lean{Algorithm} type alone does not justify stopping that review.

\paragraph{Lines.}
The program records the source range of each reached declaration when available; compiler-generated auxiliary constants either have no range or share the range of the declaration that produced them.
For each module the ranges of the reached declarations of the kinds above (theorem leaves excluded) are merged, and the non-blank lines inside them are counted after line, block and documentation comments have been removed, so a declaration is counted once however many of its constants are reached, and a documentation comment costs nothing.
A constant whose type concludes with \lean{Algorithm} contributes the lines of its signature, up to the line that holds \lean{:=} or \lean{where}.
The statement of \lean{main} is counted separately: the lines of its declaration up to and including the line on which its proof begins.
LOC is the number of non-blank, non-comment lines in all modules of the import closure of the module exporting \lean{main} inside the library, rebuilt from its import commands, including \lean{public import}, \lean{import all} and commands spanning several lines; Mathlib and the Lean core are not counted.
In the order of the rows of Table~\ref{tab:footprint}, these import closures contain 246, 244, 316, 299, 409, 492, 675, 1\,468 and 1\,507 modules.

\begin{figure}[t]
  \centering
  \includegraphics[width=0.9\linewidth]{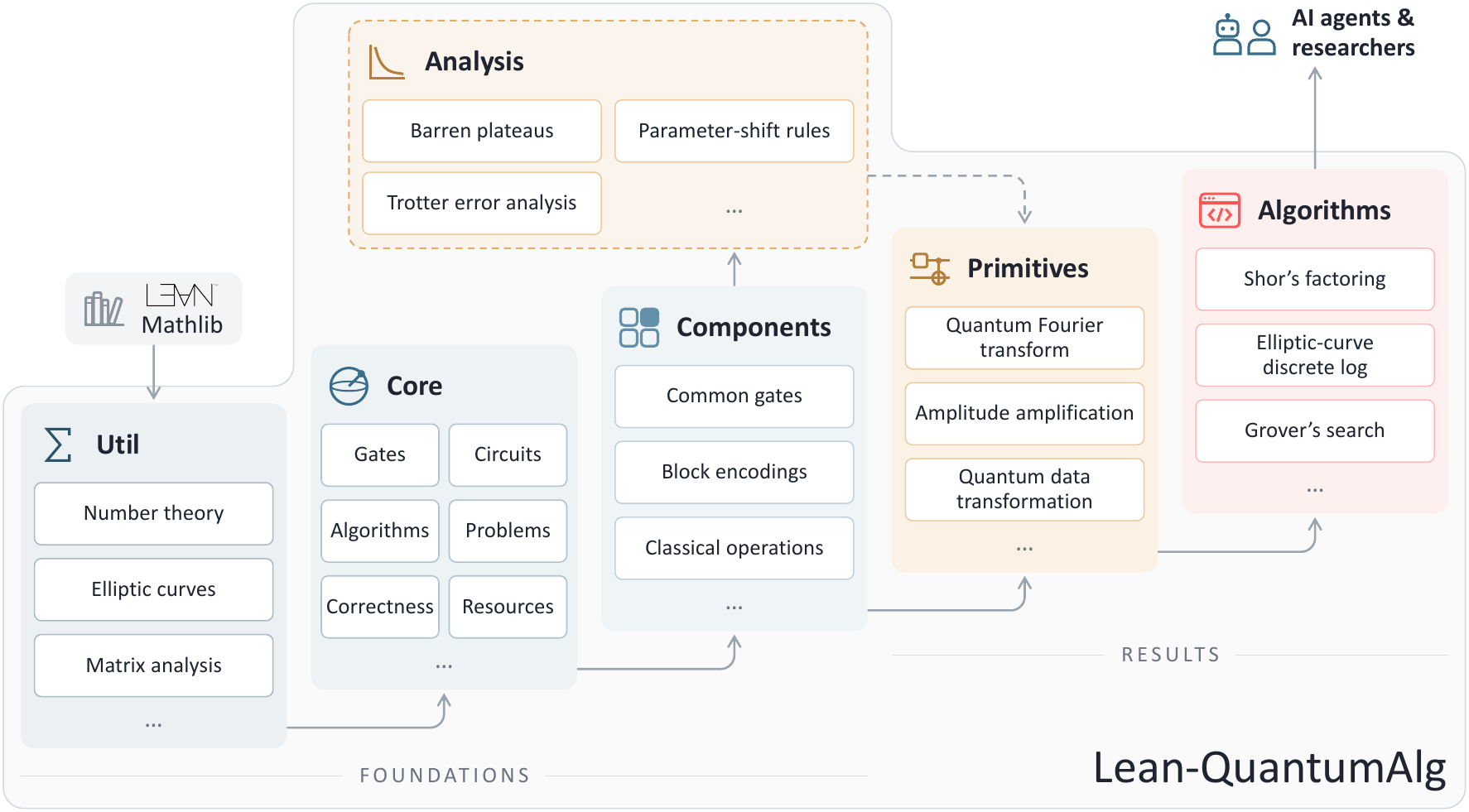}
  \caption{The module structure of \LeanQuantumAlg{}, built on Mathlib.}
  \label{fig:architecture}
\end{figure}

\paragraph{Trusted modules and theorem counts.}
The three percentages exclude different groups of the modules in Figure~\ref{fig:architecture}, identified by module prefix, and divide the remaining line counts by LOC.
For the column ``Framework untrusted'', every counted line is read.
For ``Core and Util trusted'', the lines of modules under \leanpath{QuantumAlg.Init}, \leanpath{QuantumAlg.Util} and \leanpath{QuantumAlg.Core} are subtracted.
For ``Framework trusted'', the lines of modules under \leanpath{QuantumAlg.Components} and \leanpath{QuantumAlg.Primitives} are subtracted as well, so that only the algorithm's own modules remain; the statement of \lean{main} is read in all three columns.
The import closures of the nine algorithms contain no module under \leanpath{QuantumAlg.Analysis}.
The theorem column is computed by a second walk that starts from \lean{main} and follows every constant, proofs included, within the library; it reports the number of theorem-kind constants reached, excluding \lean{main} itself and compiler-generated auxiliaries.
Declarations written with either \lean{theorem} or \lean{lemma} contribute to this count.

\paragraph{Limitations.}
A budget written as a function of the problem's parameters expands into the count together with every definition its body mentions.
This is intended, since a reviewer who wants to know what the budget says has to read those definitions, but it makes an algorithm with an arithmetic-heavy budget look larger than one whose budget is a literal.
Every counted line is weighted equally, although a budget formula and a routine field declaration cost a reviewer very different amounts of attention, and the statement of \lean{main} is counted by its lines although it is the one piece a reviewer reads most carefully.
The denominator includes all library code in the import closure, even when the final theorem does not depend on it; adding unused imports or code can therefore lower a percentage without reducing the definitions to review.
Declarations without source ranges do not contribute lines, and any proof text on the line containing a declaration's proof delimiter follows the same line-counting convention.
Import closures overlap across algorithms, so their line and theorem counts cannot be added to obtain unique library totals.
The measure identifies a source-level scope for review; it does not measure review time, understanding or the ability to detect a mistaken specification.
\section{Further algorithm examples}
\label{appendix:cases}

The quantum phase estimation example in Section~\ref{sec:framework} exercises the promise, the acceptance relation, a non-trivial threshold and resource bounds.
The algorithms below give further examples of an oracle granted by the problem, a state requirement on a construction with mid-circuit measurement, and parameters that tie validity, acceptance, threshold and budgets together.
The two displayed Lean declarations state the claims discussed below; their proofs are omitted.

\subsection{Bernstein--Vazirani: an oracle granted by the problem}
\label{appendix:bv}

For Bernstein--Vazirani~\cite{bernstein1997quantum}, the problem's \lean{Witness} is a hidden string $s \in \{0,1\}^n$, and \lean{valid} requires the environment's quantum oracle to interpret that string as the granted inner-product oracle.
The input context is one classical cell of unit shape and no quantum system, so it contains no classical copy of $s$.
Once the environment and witness agree in this way, every local input is admitted because the program prepares its own qubits; the same program is certified for every hidden string at a fixed width.
Acceptance is exact, since the returned $n$ bits must be the big-endian encoding of $s$, and the threshold is one.

The oracle query is an event of the model and is bounded by the budget of the problem like a gate: the budget allows one query, $2n + 2$ single-qubit gates, $n + 1$ qubit preparations, $n$ measurements and one bit-packing operation.
The budgets hold for every width $n$, including $n = 0$, because the theorem is quantified over $n$ and makes no claim about sample widths only.

\subsection{Measured uncomputation: a state requirement and mid-circuit measurement}
\label{appendix:mu}

In-place modular multiplication by measured uncomputation is a primitive whose result is a quantum state, not a classical answer.
Its problem has a trivial acceptance relation and threshold one; its functional requirement appears only in \lean{states}.
For a prime modulus $p$, the input is any state of two owned words $X$ and $Y$ supported on $0 < X < p$ and $0 \le Y < p$.
After all measurement records are summed out, that requirement compares the joint state of all output quantum systems and the external reference with the input state after replacing $Y$ by $X \cdot Y \bmod p$, returning the auxiliary register to the fixed value of its preparation, and retaining every correlation with the reference.

\begin{leanlisting}{Measured in-place multiplication: algorithm and final theorem (proof omitted)}{lst:mu-main}
-- source: QuantumAlg/Primitives/MeasuredUncompute/InPlace/Multiplication/Endpoint.lean:107-116
public def algorithm (modulus : Nat) (prime : Nat.Prime modulus) :
    Algorithm (.quantum AlgorithmGate.model .empty)
      (inputContext modulus.size) (outputContext modulus.size)
      (.fixedVector modulus.size (.finite 2)) where
  run := { code := code modulus prime, output := output modulus.size }

public theorem main (modulus : Nat) (prime : Nat.Prime modulus) :
    ∃ candidate, Algorithm.Solves (allowAllocation := true) candidate (problem modulus prime) := by
\end{leanlisting}

The construction measures a register in the middle of the circuit and applies a correction selected by the outcome.
Under the framework the measurement records are live classical cells declared in the final storage, and the correction is control that depends on their values.
Quantum time follows that dependency and includes the quantum gates of the selected correction; the measurement and classical processing have zero quantum duration and remain in the event budget.
The resource obligations range over execution paths with positive probability, so every correction branch that can occur is bounded.

Both input words are owned systems, so the restoration obligation \lean{borrowed} holds trivially here.
The requirement is joint: checking the target-word marginal alone could miss an incorrect auxiliary state or altered correlations with the external reference, whereas the specified joint equality constrains both.

\subsection{Amplitude estimation: parameters that tie the fields together}
\label{appendix:ae}

For amplitude estimation~\cite{brassard2002quantum}, the problem's \lean{Witness} supplies the state preparation and designated good state, and \lean{valid} binds the environment's quantum oracle to the interpretation determined by them.
Its additive error $\varepsilon$ and failure parameter $\eta$ select the number of phase qubits $t = \lceil \log_2 (\pi / (\varepsilon\eta)) \rceil$.
The valid range is $0 < \varepsilon \le 1$ and $0 < \eta \le 1$; parameters outside it admit no inputs.
The lower bound is $1 - \eta$, which is positive for $0 < \eta < 1$; $\eta = 1$ is admitted and gives the lower bound zero.
Acceptance is the additive-error test: the returned bits are decoded in big-endian order as $k$, and $\sin^2(\pi k / 2^t)$ must lie within $\varepsilon$ of the probability $p$ that the prepared state is found in the designated good state.
Every resource bound is a function of $t$ and of the number $n$ of qubits on which the state preparation acts.

\begin{leanlisting}{Amplitude estimation: promise, acceptance and final theorem (proof omitted)}{lst:ae-main}
-- source: QuantumAlg/Algorithms/AmplitudeEstimation/Endpoint/Results.lean:57-74
public lemma problem_valid_iff {n : Nat} (E : GeneralAmplitudeEstimationInput n)
    (eps η : Real) (nonnegative : 0 ≤ η) (environment : (model n).Environment)
    (state : ExecutionState (Execution.Store pre) pre.register) :
    (problem n eps η nonnegative).valid environment E state ↔
      environment.quantum = AmplitudeAmplificationSource.interpretation
        E.source.preparation E.source.good true true ∧ 0 < eps ∧ eps ≤ 1 ∧ 0 < η ∧ η ≤ 1 := Iff.rfl

public lemma problem_accepts_bits_iff {n : Nat} (E : GeneralAmplitudeEstimationInput n)
    (eps η : Real) (nonnegative : 0 ≤ η) (environment : (model n).Environment)
    (initial : Execution.Store pre)
    (outcome : Fin (selectedPhaseQubits eps η) → Bool) :
    (problem n eps η nonnegative).accepts environment E initial (.bits outcome) ↔
      |Real.sin (Real.pi *
        ((TerminalZMeasure.fullRegisterIndex outcome).val /
          (2 : Real) ^ selectedPhaseQubits eps η)) ^ 2 - targetProbability E.source| ≤ eps :=
  Iff.rfl

-- source: QuantumAlg/Algorithms/AmplitudeEstimation/Endpoint.lean:144-149
public theorem main (n : Nat) (eps η : Real) (nonnegative : 0 ≤ η) :
    ∃ candidate : Algorithm (model n) pre (post (selectedPhaseQubits eps η) n)
        (.bits (selectedPhaseQubits eps η)),
      Algorithm.Solves candidate (problem n eps η nonnegative) :=
\end{leanlisting}

For fixed $n$, $\varepsilon$ and $\eta$, one candidate in \lean{main} applies to all admitted source instances.
The same parameters appear in \lean{valid}, \lean{accepts}, \lean{successLB} and the budgets, making their relation explicit in one specification.
The selected width also fixes the output context and the output shape of the problem.
Changing that width therefore changes both the represented answer and the resource limits against which an implementation is certified; the reviewer can check these consequences together.

Separate results accompanying the endpoint state exact counts at the level of the alphabet of the model: with $L = 2^t - 1$, every path of the exported algorithm makes $L + 1$ queries to the state preparation, $L$ queries to its inverse and $L$ controlled reflections about the good state, and the number of elementary gates is given in closed form.
These theorems concern the paths of the same algorithm exhibited by the proof of \lean{main}.

\subsection{Quantum attacks on RSA-2048 and P-256}
\label{appendix:cryptographic}

\paragraph{RSA-2048.}
The short-discrete-logarithm route for factoring is described in Ref.~\cite{ekera2017quantum,ekera2023shortdlp}.
The RSA-2048 theorem applies to a 2048-bit integer $N=pq$, where $p$ and $q$ are distinct odd 1024-bit primes and $\gcd(p-1,q-1)<2^{181}$; the program receives only $N$.
Under these conditions, it returns one of the factors $p,q$ with probability at least $0.99$.
The complete run has seven batches and 112 measured samples.
It uses native gates and classical operations, with no quantum oracle.
Its bounds cover gate events, quantum time, space and classical operations, including preparation, classical processing and final selection.

\paragraph{P-256.}
The elliptic-curve discrete-logarithm method and P-256 parameters are described in Ref.~\cite{proos2004shor,roetteler2017quantum,nist2023sp800186}.
For the certified P-256 curve, a given base point of known prime order and a given point in its subgroup, the program returns a number in the subgroup range whose multiple of the base point is the given point, with probability at least $0.99$.
The unknown number is not an input.
The complete run makes 4,714 repetitions, checks each recovered number and retains the first accepted result.
Its bounds cover gate events, quantum time, space and classical operations, including preparation, reset, sampling, classical processing and final packing, with no quantum oracle.
\section{Comparing formal statements with intended claims}
\label{appendix:drift}

A formal statement can have a valid proof while expressing a claim different from the one intended by its source.
The following examples explain consequences of the definitions in Section~\ref{sec:framework} and Section~\ref{sec:execution} and the human review described in Section~\ref{sec:review}.
The examples follow from the definitions; they are not results of experiments with agents.

First, a development may narrow \lean{valid} or add a hypothesis to \lean{main} to avoid inputs on which the intended claim is difficult to prove.
For example, a promise restricted to the all-zero string changes a task intended for every $n$-bit input.
Setting \lean{accepts} to \lean{True} likewise admits every returned value, so the success obligation no longer establishes that the intended answer was returned.
The resulting theorem may be valid, so this change must be judged by comparing the specification and theorem with the source.
The framework makes that comparison local: \lean{Solves} quantifies over the instances admitted by the named problem, and a restriction used by an intermediate lemma must be discharged for those instances before it can support the final theorem.

By contrast, the framework itself prevents a gate count for one circuit from discharging the resource obligation of another.
All resource fields of \lean{Solves} range over execution paths with positive probability in the algorithm's circuit, whose execution also determines the success probability and final state.
Thus, proving that an optimized variant has a small gate count provides no certificate for the original circuit unless the required bound is also established for that circuit.
The shared run enforces this connection independently of how the proofs are organized.

Relabeling a composite operation as a native gate has a different boundary.
Within a fixed execution model, a unitary leaf must use a declared token whose event coordinate is fixed; a gate's operator is fixed by its token, while an oracle query's operator is selected by the environment's interpretation of its declared oracle.
A model that grants SWAP as one native gate counts that operation once, although the implementation in Figure~\ref{fig:swap} uses three CNOT gates.
Choosing that model changes an assumption the reviewer reads, including the meaning of the resource unit.
The same reasoning applies to granting a new oracle: the resulting certificate is relative to that declaration, and the reviewer must judge whether the source grants the same access.

The budget itself also requires review.
Suppose a circuit contains $1\,000$ CNOT gates and the intended claim permits at most $100$.
If the problem's CNOT budget is defined as that circuit's own CNOT count, its resource obligation becomes $1\,000 \le 1\,000$.
The comparison correctly uses the actual circuit, but the chosen budget has changed the resource claim.
The reviewer therefore follows the budget's defining expressions and compares them with the intended limit, while the fixed \lean{Solves} relation checks the implementation against the requirement that was stated.

Borrowed-state restoration illustrates why the specification must also account for correlations.
Suppose a borrowed qubit is half of a Bell pair with an external reference.
Returning the same maximally mixed marginal while destroying its correlation with the reference would pass a check of that marginal alone.
This change fails \lean{RestoresBorrowed}, which compares the joint borrowed system and reference before and after execution.
Since this is a separate field of \lean{Solves}, satisfying a weaker output-state requirement cannot replace it.

Finally, classical work represented by the run contributes events and is subject to the corresponding budget, even though its canonical quantum duration is zero.
An implementation cannot erase such an event from the resource account while retaining that operation in the run.
Work performed outside the formalized execution is outside the certificate as well.
Determining whether the source algorithm requires additional preparation or post-processing therefore remains part of reviewing the specification and model; the framework cannot infer omitted external work from a proof of the formalized claim.

\begin{thebibliography}{10}

\bibitem{grover1996fast}
Lov~K. Grover.
\newblock A fast quantum mechanical algorithm for database search.
\newblock In {\em Proceedings of the Twenty-Eighth Annual {ACM} Symposium on
  Theory of Computing ({STOC} '96)}, pages 212--219, New York, 1996. ACM Press.

\bibitem{shor1997polynomial}
Peter~W. Shor.
\newblock Polynomial-time algorithms for prime factorization and discrete
  logarithms on a quantum computer.
\newblock {\em {SIAM} Journal on Computing}, 26(5):1484--1509, 1997.

\bibitem{ruiz2025alphatensorq}
Francisco J.~R. Ruiz, Tuomas Laakkonen, Johannes Bausch, Matej Balog,
  Mohammadamin Barekatain, Francisco J.~H. Heras, Alexander Novikov, Nathan
  Fitzpatrick, Bernardino Romera-Paredes, John van~de Wetering, Alhussein
  Fawzi, Konstantinos Meichanetzidis, and Pushmeet Kohli.
\newblock Quantum circuit optimization with {AlphaTensor}.
\newblock {\em Nature Machine Intelligence}, 7(3):374--385, 2025.

\bibitem{deepmind2026alphaevolve}
{Google DeepMind}.
\newblock {AlphaEvolve}: How our {Gemini}-powered coding agent is scaling
  impact across fields.
\newblock \url{https://deepmind.google/blog/alphaevolve-impact/}, 2026.
\newblock accessed October 2026.

\bibitem{tsoukalas2026nexus}
George Tsoukalas, Anton Kovsharov, Sergey Shirobokov, Anja Surina, Moritz
  Firsching, Gergely B{\'e}rczi, Francisco J.~R. Ruiz, Arun Suggala, Adam~Zsolt
  Wagner, Eric Wieser, Lei Yu, Aja Huang, Mikl{\'o}s~Z. Horv{\'a}th, Andrew
  Ferraiuolo, Henryk Michalewski, Codrut Grosu, Edward Lockhart, Thomas Hubert,
  Matej Balog, Pushmeet Kohli, and Swarat Chaudhuri.
\newblock Advancing mathematics research with {AI}-driven formal proof search.
\newblock {\em Science}, 394(6820):234--239, 2026.

\bibitem{demoura2021lean}
Leonardo de~Moura and Sebastian Ullrich.
\newblock The {Lean} 4 theorem prover and programming language.
\newblock In {\em Automated Deduction -- {CADE} 28}, volume 12699 of {\em
  Lecture Notes in Computer Science}, pages 625--635, 2021.

\bibitem{openai2026navierstokes}
{OpenAI}.
\newblock On the {Navier--Stokes} {Millennium Prize Problem}.
\newblock \url{https://openai.com/index/navier-stokes-solution/}, 2026.
\newblock accessed October 2026.

\bibitem{anthropic2026flt}
{Anthropic}.
\newblock Formalizing {Fermat's Last Theorem}.
\newblock
  \url{https://www.anthropic.com/research/formalizing-fermats-last-theorem},
  2026.
\newblock accessed October 2026.

\bibitem{openai2026mathrepo}
{OpenAI}.
\newblock {openai/math}.
\newblock \url{https://github.com/openai/math}, 2026.
\newblock accessed October 2026.

\bibitem{sothanaphan2026erdos728}
Nat Sothanaphan.
\newblock Resolution of {Erd{\H{o}}s} problem \#728: a writeup of {Aristotle}'s
  {Lean} proof.
\newblock \url{https://arxiv.org/abs/2601.07421}, 2026.

\bibitem{ammanamanchi2026faults}
Pawan~Sasanka Ammanamanchi, Siddharth Bhat, and Stella Biderman.
\newblock Faults in our formal benchmarking: Dataset defects and evaluation
  failures in {Lean} theorem proving.
\newblock In {\em Proceedings of the 43rd International Conference on Machine
  Learning}, volume 306 of {\em Proceedings of Machine Learning Research},
  pages 2431--2462, 2026.

\bibitem{bastounis2026navierstokes}
Alexander Bastounis, Fabian Circelli, and Anders~C. Hansen.
\newblock {Navier--Stokes} lost in translation: Why {Lean} verification of {AI}
  autoformalisation does not guarantee correct natural language proofs.
\newblock \url{https://arxiv.org/abs/2610.08144}, 2026.

\bibitem{tao2026ageofai}
Terence Tao.
\newblock Mathematics in the age of {AI}.
\newblock \url{https://arxiv.org/abs/2608.16753}, 2026.

\bibitem{nielsen2010quantum}
Michael~A. Nielsen and Isaac~L. Chuang.
\newblock {\em Quantum Computation and Quantum Information}.
\newblock Cambridge University Press, Cambridge, 10th anniversary edition,
  2010.

\bibitem{gidney2017factoring}
Craig Gidney.
\newblock Factoring with $n+2$ clean qubits and $n-1$ dirty qubits.
\newblock \url{https://arxiv.org/abs/1706.07884}, 2017.

\bibitem{chareton2021automated}
Christophe Chareton, S{\'e}bastien Bardin, Fran{\c{c}}ois Bobot, Valentin
  Perrelle, and Beno{\^i}t Valiron.
\newblock An automated deductive verification framework for circuit-building
  quantum programs.
\newblock In Nobuko Yoshida, editor, {\em Programming Languages and Systems},
  volume 12648 of {\em Lecture Notes in Computer Science}, pages 148--177,
  Cham, 2021. Springer International Publishing.

\bibitem{hietala2021proving}
Kesha Hietala, Robert Rand, Shih-Han Hung, Liyi Li, and Michael Hicks.
\newblock Proving quantum programs correct.
\newblock In {\em 12th International Conference on Interactive Theorem Proving
  ({ITP} 2021)}, volume 193 of {\em Leibniz International Proceedings in
  Informatics ({LIPIcs})}, pages 21:1--21:19. Schloss Dagstuhl --
  Leibniz-Zentrum f{\"u}r Informatik, 2021.

\bibitem{rand2018qwire}
Robert Rand, Jennifer Paykin, and Steve Zdancewic.
\newblock {QWIRE} practice: Formal verification of quantum circuits in {Coq}.
\newblock {\em Electronic Proceedings in Theoretical Computer Science},
  266:119--132, 2018.

\bibitem{zhou2023coqq}
Li~Zhou, Gilles Barthe, Pierre-Yves Strub, Junyi Liu, and Mingsheng Ying.
\newblock {CoqQ}: Foundational verification of quantum programs.
\newblock {\em Proceedings of the {ACM} on Programming Languages},
  7(POPL):833--865, 2023.

\bibitem{ren2026qedia}
Yuanjie~(Collin) Ren, Zhicheng Zhang, Junyi Liu, Jinzheng Li, Yidi Qi, and
  Mingsheng Ying.
\newblock {QEDIA}: Quantum program verification in {Lean}.
\newblock \url{https://arxiv.org/abs/2610.17708v1}, 2026.

\bibitem{peng2023formally}
Yuxiang Peng, Kesha Hietala, Runzhou Tao, Liyi Li, Robert Rand, Michael Hicks,
  and Xiaodi Wu.
\newblock A formally certified end-to-end implementation of {Shor}'s
  factorization algorithm.
\newblock {\em Proceedings of the National Academy of Sciences},
  120(21):e2218775120, May 2023.

\bibitem{colledan2025circuit}
Andrea Colledan, Ugo Dal~Lago, and Niki Vazou.
\newblock Circuit width estimation via effect typing and linear dependency.
\newblock {\em ACM Transactions on Programming Languages and Systems},
  47(3):13:1--13:35, 2025.

\bibitem{colledan2025flexible}
Andrea Colledan and Ugo Dal~Lago.
\newblock Flexible type-based resource estimation in quantum circuit
  description languages.
\newblock {\em Proceedings of the {ACM} on Programming Languages},
  9(POPL):47:1--47:31, 2025.

\bibitem{green2013quipper}
Alexander~S. Green, Peter~LeFanu Lumsdaine, Neil~J. Ross, Peter Selinger, and
  Beno{\^i}t Valiron.
\newblock Quipper: A scalable quantum programming language.
\newblock In {\em Proceedings of the 34th {ACM} {SIGPLAN} Conference on
  Programming Language Design and Implementation}, pages 333--342, 2013.

\bibitem{harrigan2024qualtran}
Matthew~P. Harrigan, Tanuj Khattar, Charles Yuan, Anurudh Peduri, Noureldin
  Yosri, Fionn~D. Malone, Ryan Babbush, and Nicholas~C. Rubin.
\newblock Expressing and analyzing quantum algorithms with {Qualtran}.
\newblock \url{https://arxiv.org/abs/2409.04643}, 2024.

\bibitem{lu2025formalalign}
Jianqiao Lu, Yingjia Wan, Yinya Huang, Jing Xiong, Zhengying Liu, and Zhijiang
  Guo.
\newblock {FormalAlign}: Automated alignment evaluation for autoformalization.
\newblock In {\em The Thirteenth International Conference on Learning
  Representations}, pages 102122--102149, 2025.
\newblock \url{https://openreview.net/forum?id=B5RrIFMqbe}.

\bibitem{zhang2026beyond}
Ke~Zhang, Patricio Gallardo~Candela, Sudhir Murthy, Yi~Xie, Zhi Wang, and
  Maziar Raissi.
\newblock Beyond compilation: Evaluating faithful natural-language-to-{Lean}
  statement formalization.
\newblock \url{https://arxiv.org/abs/2606.31002}, 2026.

\bibitem{pegden2026trellis}
Wesley Pegden.
\newblock ({A}uto)formalization is supposed to be easy: Trellis process
  semantics for spelling out rigorous proofs.
\newblock \url{https://arxiv.org/abs/2606.09674}, 2026.

\bibitem{mathlib}
{The mathlib Community}.
\newblock The {Lean} mathematical library.
\newblock In {\em Proceedings of the 9th {ACM} {SIGPLAN} International
  Conference on Certified Programs and Proofs}, CPP 2020, pages 367--381.
  Association for Computing Machinery, 2020.

\bibitem{deutsch1992rapid}
David Deutsch and Richard Jozsa.
\newblock Rapid solution of problems by quantum computation.
\newblock {\em Proceedings of the Royal Society of London. Series A:
  Mathematical and Physical Sciences}, 439(1907):553--558, 1992.

\bibitem{dewolf2019lecturenotes}
Ronald de~Wolf.
\newblock Quantum computing: Lecture notes, 2019.

\bibitem{bernstein1997quantum}
Ethan Bernstein and Umesh Vazirani.
\newblock Quantum complexity theory.
\newblock {\em {SIAM} Journal on Computing}, 26(5):1411--1473, 1997.

\bibitem{kitaev1995quantum}
A.~Yu. Kitaev.
\newblock Quantum measurements and the {A}belian stabilizer problem, 1995.

\bibitem{lin2022qasc}
Lin Lin.
\newblock Lecture notes on quantum algorithms for scientific computation.
\newblock \url{https://arxiv.org/abs/2201.08309}, 2022.

\bibitem{brassard2002quantum}
Gilles Brassard, Peter H{\o}yer, Michele Mosca, and Alain Tapp.
\newblock Quantum amplitude amplification and estimation.
\newblock {\em Contemporary Mathematics}, 305:53--74, 2002.

\bibitem{shukla2024uniformsuperposition}
Alok Shukla and Prakash Vedula.
\newblock An efficient quantum algorithm for preparation of uniform quantum
  superposition states.
\newblock {\em Quantum Information Processing}, 23(2):38, 2024.

\bibitem{ekera2017quantum}
Martin Eker{\aa} and Johan H{\aa}stad.
\newblock Quantum algorithms for computing short discrete logarithms and
  factoring {RSA} integers.
\newblock In {\em Post-Quantum Cryptography}, volume 10346 of {\em Lecture
  Notes in Computer Science}, pages 347--363. Springer, 2017.

\bibitem{ekera2023shortdlp}
Martin Eker{\aa}.
\newblock On the success probability of the quantum algorithm for the short
  {DLP}.
\newblock {\em {IACR} Communications in Cryptology}, 3(1):10, 2026.
\newblock \url{https://cic.iacr.org/p/3/1/10}.

\bibitem{proos2004shor}
John Proos and Christof Zalka.
\newblock {Shor}'s discrete logarithm quantum algorithm for elliptic curves.
\newblock {\em Quantum Information and Computation}, 3(4):317--344, 2003.

\bibitem{roetteler2017quantum}
Martin Roetteler, Michael Naehrig, Krysta~M. Svore, and Kristin Lauter.
\newblock Quantum resource estimates for computing elliptic curve discrete
  logarithms.
\newblock In {\em Advances in Cryptology -- {ASIACRYPT} 2017}, volume 10625 of
  {\em Lecture Notes in Computer Science}, pages 241--270. Springer, 2017.

\bibitem{nist2023sp800186}
Lily Chen, Dustin Moody, Karen Randall, Andrew Regenscheid, and Angela
  Robinson.
\newblock Recommendations for discrete logarithm-based cryptography: Elliptic
  curve domain parameters.
\newblock Technical Report NIST Special Publication 800-186, National Institute
  of Standards and Technology, 2023.

\bibitem{hietala2021verified}
Kesha Hietala, Robert Rand, Shih-Han Hung, Xiaodi Wu, and Michael Hicks.
\newblock A verified optimizer for quantum circuits.
\newblock {\em Proceedings of the {ACM} on Programming Languages},
  5(POPL):37:1--37:29, 2021.

\bibitem{paykin2017qwire}
Jennifer Paykin, Robert Rand, and Steve Zdancewic.
\newblock {QWIRE}: a core language for quantum circuits.
\newblock In {\em Proceedings of the 44th {ACM} {SIGPLAN} Symposium on
  Principles of Programming Languages}, pages 846--858, 2017.

\bibitem{lean2026validating}
{Lean Language Reference}.
\newblock Validating a {Lean} proof.
\newblock \url{https://lean-lang.org/doc/reference/latest/ValidatingProofs/},
  2026.
\newblock {Lean} version used: 4.33.1; accessed October 2026.

\end{thebibliography}
\end{document}